\documentclass[12pt]{article}

\usepackage{axodraw}

\begin{document}
\begin{titlepage}
\begin{flushright}
UFIFT-QG-05-07 \\ gr-qc/0508015
\end{flushright}
\vspace{.4cm}
\begin{center}
\textbf{Charged Scalar Self-Mass during Inflation}
\end{center}
\begin{center}
E. O. Kahya$^{\dagger}$ and R. P. Woodard$^{\ddagger}$
\end{center}
\begin{center}
\textit{Department of Physics \\ University of Florida \\
Gainesville, FL 32611 USA}
\end{center}

\begin{center}
ABSTRACT
\end{center}
We compute the one loop self-mass of a charged massless, minimally coupled
scalar in a locally de Sitter background geometry. The computation is done
in two different gauges: the noninvariant generalization of Feynman gauge
which gives the simplest expression for the photon propagator and the de
Sitter invariant gauge of Allen and Jacobson. In each case dimensional
regularization is employed and fully renormalized results are obtained. 
By using our result in the linearized, effective field equations one can
infer how the scalar responds to the dielectric medium produced by 
inflationary particle production. We also work out the result for
a conformally coupled scalar. Although the conformally coupled case is of 
no great physical interest the fact that we obtain a manifestly de Sitter 
invariant form for its self-mass-squared establishes that our noninvariant 
gauge introduces no physical breaking of de Sitter invariance at one loop 
order.

\begin{flushleft}
\end{flushleft}
\vspace{.4cm}
\begin{flushleft}
$^{\dagger}$ e-mail: emre@phys.ufl.edu \\
$^{\ddagger}$ e-mail: woodard@phys.ufl.edu
\end{flushleft}
\end{titlepage}

\section{Introduction}

Quantum loop effects can be understood in very simple terms as the response
of classical field theory to the 0-point fluctuations of the various dynamical
variables. These fluctuations can be viewed as virtual particles which emerge
from the vacuum and disappear back into it after a period whose duration is
fixed by the energy-time uncertainty principle. Because spacetime expansion
redshifts virtual particle energies it increases the time for which they can
persist, hence strengthening the effects they produce. Parker was the first
to perform serious computations about this \cite{LP}.

Just as in flat space, the persistence time for virtual particles of the same
momentum becomes longer as the mass decreases. It is simple to show that any
massless and sufficiently long wavelength virtual particle which happens to
emerge from the vacuum can persist forever during inflation \cite{PW1,RW1,RW2}.
However, this will not lead to increased quantum effects unless the rate at
which such particles emerge from the vacuum is significant. Almost all
masslessless particles possess classical conformal invariance, which causes
the rate at which they emergence from the vacuum to redshift so rapidly that
there is no significant strengthening of quantum effects \cite{PW1,RW1,RW2}. 
The two exceptions are massless, minimally coupled (MMC) scalars and gravitons.
These particles are produced copiously during inflation and can therefore
mediate enhanced quantum effects.

It is quantum fluctuations of precisely these fields which are responsible
for the primordial cosmological scalar \cite{Slava} and tensor \cite{Alexei}
perturbations predicted by inflation \cite{MFB,LL}. Those are tree order
effects. At one loop order it has recently been discovered that MMC scalars
can catalyze significant quantum effects involving particles which would not
otherwise experience them owing to classical conformal invariance. When
electromagnetism (which is conformally invariant for $D\!=\!4$ dimensions) is
coupled to a charged MMC scalar the resulting one loop vacuum polarization
causes super-horizon photons to behave, in some ways, as though they have
nonzero mass \cite{PTW1,PTW2,PW2}. When massless fermions (which are
conformally invariant in any dimension) are Yukawa coupled to a MMC scalar
the resulting one loop fermion self-energy engenders a faster-than-exponential
growth in the fermion mode functions which seems to betoken explosive particle
production \cite{PW3}.

Both models have great phenomenological interest because all Standard Model
particles would be effectively massless on the energy scales typically
envisaged for primordial inflation, and because they couple to a fundamental
scalar --- the Higgs --- which may well be minimally coupled. If unchecked,
the Yukawa process would lead to a degenerate fermi gas of super-horizon
fermions \cite{PW3}. The electromagnetic process would not result in photon 
creation during inflation but leads instead to a vast enhancement of the 
0-point energy of super-horizon photons. After the end of inflation the 
least super-horizon of these modes will re-enter the horizon and again 
become massless, at which point their excess 0-point energies may seed 
cosmic magnetic fields \cite{DDPT,DPTD,PW1}.

Both effects require the scalar to be massless on the scale of the
inflationary Hubble constant and it is natural to wonder whether quantum
corrections can affect this. What happens can be determined by computing 
the self-mass-squared and using this to solve the linearized effective 
field equations,
\begin{equation}
\partial_{\mu} \Bigl(\sqrt{-g} g^{\mu\nu} \partial_{\nu} \varphi \Bigr)
- \int d^4x' M^2(x;x') \varphi(x') = 0 \; . \label{effeqn}
\end{equation}
In a previous paper it was shown that the scalar cannot acquire a large 
enough mass rapidly enough to prevent the Yukawa process from going to 
completion \cite{DW}. That result was inevitable because only the 
conformally invariant fermion propagator contributes to the Yukawa 
self-mass-squared at one loop order. The non-conformal scalar propagator 
participates even at one loop order in SQED so there should be a strong
effect. Because inflation induces an effective dielectric medium of
super-horizon charged scalars \cite{PW1}, and because the energy of a 
charged particle is reduced in a dielectric medium, it is conceivable 
that the effect will be to {\it enhance} inflationary particle production. 
In this paper we evaluate $M^2(x;x')$ at one loop; we will use it in a 
subsequent paper to solve the effective field equation (\ref{effeqn}).

Because this computation involves the photon propagator we must face the
issue of gauge fixing. It turns out that the simplest gauge breaks de Sitter
invariance \cite{RPW}. Because the scalar propagator must show a physical
--- as opposed to gauge --- breaking of de Sitter invariance \cite{AF} this
should not be an issue. However, one might worry about the introduction of
spurious violations of de Sitter invariance. That can be checked by
computing the self-mass-squared of the conformally coupled scalar, which 
shows no physical breaking of de Sitter invariance and for which we obtain
a manifestly de Sitter invariant result. Another check is by computing the
minimally coupled scalar self-mass-squared in the de Sitter invariant gauge 
of Allen and Jacobson \cite{AJ}.

In the next section we use the Lagrangian of SQED to derive the Feynman
rules in the simplest gauge. In section 3 we compute the fully renormalized
one loop scalar self-mass-squared. In section 4 the same quantity is computed
in the de Sitter invariant gauge of Allen and Jacobson. What it all means is 
discussed in section 5.

\section{Feynman Rules}

Let $\varphi(x)$ represent a complex scalar field and let $A_{\mu}(x)$ stand
for the vector potential. The Lagrangian of MMC scalar QED is,
\begin{eqnarray}
\lefteqn{\mathcal{L}_{\mbox{\tiny SQED}} = - \Bigl(\partial_{\mu} \!-\! i e_0
A_{\mu}\Bigr) \varphi^* \Bigl(\partial_{\nu} \!+\! ie_0 A_{\nu}\Bigr) \varphi
\, g^{\mu\nu} \sqrt{-g} } \nonumber \\
& & \hspace{4cm} - \xi_0 \varphi^* \varphi R \sqrt{-g} - \frac14 F_{\mu\nu}
F_{\rho\sigma} g^{\mu\rho} g^{\nu\sigma} \sqrt{-g} \; . \qquad
\end{eqnarray}
Here $\xi_0$ is the bare conformal coupling and $e_0$ is the bare charge.
We do not need to include a bare mass because mass is multiplicatively
renormalized in dimensional regularization. We {\it would} need a $(\varphi^*
\varphi)^2$ interaction for strict renormalizability but we shall not require
such a term for this computation.

Renormalization is organized in the usual way. We first define renormalized
fields in terms of the bare ones,
\begin{equation}
\varphi \equiv \sqrt{Z}_2 \varphi_r \qquad {\rm and} \qquad A_{\mu} \equiv
\sqrt{Z}_3 A_{r\mu} \; .
\end{equation}
In terms of the renormalized fields the Lagrangian takes the form,
\begin{eqnarray}
\lefteqn{\mathcal{L}_{\mbox{\tiny SQED}} = - Z_2 \Bigl(\partial_{\mu} \!-\!
i e_0 \sqrt{Z}_3 A_{r \mu}\Bigr) \varphi_r^* i \Bigl(\partial_{\nu} \!+\! i e_0
\sqrt{Z}_3 A_{r \nu}\Bigr) \varphi_r g^{\mu\nu} \sqrt{-g} } \nonumber \\
& & \hspace{3.5cm} - \xi_0 Z_2 \varphi_r^* \varphi_r R \sqrt{-g} - \frac14 Z_3
F_{r \mu\nu} F_{r \rho\sigma} g^{\mu\rho} g^{\nu\sigma} \sqrt{-g} \; . \qquad
\end{eqnarray}
We will henceforth speak only of the renormalized fields and, because there
can be no more ambiguity with unrenormalized fields, we will dispense with the 
superfluous subscript ``r''. Because of gauge invariance and the fact that we 
want the tree order scalar to be minimally coupled our renormalized parameters 
are given by the relations,
\begin{equation}
\sqrt{Z}_3 e_0 = e + 0 \qquad {\rm and} \qquad Z_2 \xi_0 = 0 + \delta \xi \; .
\end{equation}
Setting $Z_{2,3} = 1 + \delta Z_{2,3}$, we finally express the Lagrangian
as,
\begin{eqnarray}
\lefteqn{\mathcal{L}_{\mbox{\tiny SQED}} = - \Bigl(\partial_{\mu} -i e
A_{\mu}\Bigr) \varphi^* \Bigl(\partial_{\nu} + i e A_{\nu}\Bigr) \varphi \,
g^{\mu\nu} \sqrt{-g} - \frac14 F_{\mu\nu} F_{\rho\sigma} g^{\mu\rho}
g^{\nu\sigma} \sqrt{-g} } \nonumber \\
& & \hspace{2.5cm} - \delta Z_2 \Bigl(\partial_{\mu} \!-\! i e A_{\mu}\Bigr)
\varphi^* \Bigl(\partial_{\nu} \!+\! i e A_{\nu}\Bigr) \varphi g^{\mu\nu}
\sqrt{-g} - \delta \xi \varphi^* \varphi R \sqrt{-g} \nonumber \\
& & \hspace{7cm} - \frac14 \delta Z_3 F_{\mu\nu} F_{\rho\sigma} g^{\mu\rho}
g^{\nu\sigma} \sqrt{-g} \; . \qquad
\end{eqnarray}

Although we can only obtain propagators in very special geometries it is
simple enough to read off the interactions needed to compute the one loop
self-mass-squared in a general metric background. The 3-point and 4-point
interactions are,
\begin{eqnarray}
\mathcal{L}_{\mbox{\tiny 3pt}} & = & ie A_{\mu} \Bigl(\varphi^* \varphi_{,\nu}
- \varphi^*_{,\nu} \varphi\Bigr) g^{\mu\nu} \sqrt{-g} \; , \label{3pt} \\
\mathcal{L}_{\mbox{\tiny 4pt}} & = & -e^2 A_{\mu} A_{\nu} \, \varphi^* \varphi
\, g^{\mu\nu} \sqrt{-g} \; . \label{4pt}
\end{eqnarray}
The counterterms we shall need are,
\begin{equation}
\mathcal{L}_{\mbox{\tiny ctm}} = -\delta Z_2 \, \varphi^*_{,\mu} \varphi_{,\nu}
g^{\mu\nu} \sqrt{-g} - \delta \xi \varphi^* \varphi R \, \sqrt{-g} \; .
\label{ctm}
\end{equation}
The scalar propagator comes from the Lagrangian for complex Klein-Gordon
theory,
\begin{equation}
\mathcal{L}_{\mbox{\tiny CKG}} = - \varphi^*_{,\mu} \varphi_{,\nu}
g^{\mu\nu} \sqrt{-g} \; .
\end{equation}
It obeys the equation,
\begin{equation}
\partial_{\mu} \Bigl( \sqrt{-g} g^{\mu\nu} \partial_{\nu} i\Delta_A(x;x')
\Bigr) = i \delta^D(x-x') \; . \label{Aeqn}
\end{equation}
The photon propagator derives from the Lagrangian for electromagnetism,
\begin{eqnarray}
\lefteqn{\mathcal{L}_{\mbox{\tiny EM}} = - \frac14 F_{\mu\nu} F_{\rho\sigma}
g^{\mu\rho} g^{\nu\sigma} \sqrt{-g} \; , } \\
& & = -\frac12 A_{\mu ; \nu} A_{\rho ; \sigma} g^{\mu\rho} g^{\nu\sigma}
\sqrt{-g} + \frac12 A_{\mu ; \nu} A_{\sigma ; \rho} g^{\mu\rho} g^{\nu\sigma}
\sqrt{-g} \; , \\
& & = -\frac12 A_{\mu ; \nu} A_{\rho ; \sigma} g^{\mu\rho} g^{\nu\sigma}
\sqrt{-g} -\frac12 A_{\mu} A_{\nu} R^{\mu\nu} \sqrt{-g} + \frac12 A_{\mu ; \nu}
A_{\rho ; \sigma} g^{\mu\nu} g^{\rho\sigma} \sqrt{-g} \nonumber \\
& & \hspace{3cm} + \frac12 \partial_{\rho} \Bigl(A_{\mu ; \nu} A_{\sigma}
g^{\mu\rho} g^{\nu\sigma} \sqrt{-g} - A_{\sigma ; \mu} A_{\nu} g^{\mu\sigma}
g^{\nu\rho} \sqrt{-g} \Bigr) \; . \qquad \label{EM}
\end{eqnarray}

Our paradigm for the geometry of inflation is the open conformal coordinate
patch of de Sitter,
\begin{equation}
ds^2 = a^2 \Bigl(-d\eta^2 + d\vec{x} \cdot d\vec{x} \Bigr) \qquad {\rm where}
\qquad a(\eta) = -\frac1{H \eta} \; . \label{conf}
\end{equation}
Note that the conformal time $\eta$ runs from $-\infty$ to zero. In this
background the various metric-dependent quantities that appear in
$\mathcal{L}_{\mbox{\tiny SQED}}$ are,
\begin{eqnarray}
& & g^{\mu\nu} = a^{-2} \eta^{\mu\nu} \quad , \quad \sqrt{-g} = a^D \quad ,
\quad \Gamma^{\rho}_{~\mu\nu} = H a \Bigl(\delta^{\rho}_{\mu} \delta^0_{\nu}
+ \delta^{\rho}_{\nu} \delta^0_{\mu} + \delta^{\rho}_0 \eta_{\mu\nu}\Bigr)
\; , \nonumber \\
& & R = D (D\!-\!1) H^2 \qquad {\rm and} \qquad R^{\mu\nu} = (D\!-\!1) H^2
a^{-2} \eta^{\mu\nu} \; .
\end{eqnarray}
We shall henceforth raise and lower indicies with the Lorentz metric,
\begin{equation}
\partial^{\mu} \equiv \eta^{\mu\nu} \partial_{\nu} \qquad , \qquad
\partial^2 \equiv \eta^{\mu\nu} \partial_{\mu} \partial_{\nu} \qquad , \qquad
A^{\mu} \equiv \eta^{\mu\nu} A_{\nu} \; .
\end{equation}

The special properties of our locally de Sitter background can be used to
simplify the three volume terms of the electromagnetic Lagrangian (\ref{EM}).
The first becomes,
\begin{eqnarray}
\lefteqn{-\frac12 A_{\mu ; \nu} A_{\rho ; \sigma} g^{\mu\rho} g^{\nu\sigma}
\sqrt{-g} = -\frac12 a^{D-4} \Biggl\{ A_{\mu , \nu} A^{\mu , \nu} + 2 H a
A_{\mu , 0} A^{\mu} } \nonumber \\
& & \hspace{.5cm} + 2 H a A_{0 , \mu} A^{\mu} - 2 H a A^{\mu}_{~ ,\mu} A_0 -
2 H^2 a^2 A_{\mu} A^{\mu} + (D\!-\!2) H^2 a^2 A_0^2\Biggr\} \; . \qquad \\
& & = -\frac12 a^{D-4} A_{\mu ,\nu} A^{\mu ,\nu} + \frac12 (D\!-\!1) H^2
a^{D-2} A_{\mu} A^{\mu} + 2 H a^{D-3} A_0 A^{\mu}_{~ ,\mu} \nonumber \\
& & \hspace{.5cm} -\frac12 (3D\!-\!8) H^2 a^{D-2} A_0^2 -\frac12 \Bigl(H
a^{D-3} A_{\mu} A^{\mu}\Bigr)_{,0} - \Bigl(H a^{D-3} A_0 A^{\mu}\Bigr)_{,\mu}
\; . \qquad
\end{eqnarray}
The second volume term is simple by comparison,
\begin{equation}
-\frac12 A_{\mu} A_{\nu} R^{\mu\nu} \sqrt{-g} = -\frac12 (D\!-\!1) H^2
a^{D-2} A_{\mu} A^{\mu} \; .
\end{equation}
The final volume term is,
\begin{equation}
\frac12 \Bigl(A_{\mu ; \nu} g^{\mu\nu}\Bigr)^2 \!\! \sqrt{-g} \!= \!\frac12
a^{D-4} (A^{\mu}_{~,\mu})^2 - (D\!-\!2) H a^{D-3} A_0 A^{\mu}_{~,\mu} + \frac12
(D\!-\!2)^2 H^2 a^{D-2} A_0^2 .
\end{equation}
The three volume terms can be summed to give,
\begin{eqnarray}
\lefteqn{\mathcal{L}_{\mbox{\tiny EM}} - \Bigl({\rm Surface\ Term}\Bigr) =
-\frac12 a^{D-4} A_{\mu ,\nu} A^{\mu ,\nu} + \frac12 (D\!-\!4) a^{D-2}
H^2 A_0^2 } \nonumber \\
& & \hspace{6cm} + \frac12 a^{D-4} \Bigl(A^{\mu}_{~ ,\mu} - (D\!-\!4) H a A_0
\Bigr)^2 . \qquad \label{EMdS}
\end{eqnarray}

The final expression in (\ref{EMdS}) suggests that we add the gauge fixing
term \cite{RPW},
\begin{equation}
\mathcal{L}_{\mbox{\tiny GF}} = -\frac12 a^{D-4} \Bigl(A^{\mu}_{~,\mu} -
(D \!-\!4) H a A_0\Bigr)^2 \; . \label{GF}
\end{equation}
In this gauge the photon propagator obeys the equation,
\begin{equation}
\Biggl(\partial^{\sigma} a^{D-4} \partial_{\sigma} \delta^{\rho}_{\mu} -
(D\!-\!4) H^2 a^{D-2} \delta^0_{\mu} \delta^{\rho}_0 \Biggl)
i\Bigl[{}_{\rho} \Delta_{\nu}\Bigr](x;x') = \eta_{\mu\nu} \,
i\delta^D(x-x') \; .
\end{equation}
Because space and time components are treated differently it is useful to
have expressions for the purely spatial parts of the Minkowski metric and
the Kronecker delta,
\begin{equation}
\overline{\eta}_{\mu\nu} \equiv \eta_{\mu\nu} + \delta^0_{\mu} \delta^0_{\nu}
\qquad , \qquad \overline{\delta}^{\mu}_{\nu} \equiv \delta^{\mu}_{\nu}
- \delta^{\mu}_0 \delta^0_{\nu} \; . \label{bar}
\end{equation}
Making the ansatz,
\begin{equation}
i\Bigl[{}_{\mu} \Delta_{\nu}\Bigr](x;x') = a a' \overline{\eta}_{\mu\nu}
i\Delta_B(x;x') - a a' \delta^0_{\mu} \delta^0_{\nu} \, i\Delta_C(x;x') \; .
\end{equation}
we see that the propagators $i\Delta_B(x;x')$ and $i\Delta_C(x;x')$
obey the equations of scalars with various amounts of conformal coupling,
\begin{eqnarray}
\Biggl[ \partial_{\mu} \Bigl(\sqrt{-g} g^{\mu\nu} \partial_{\nu}\Bigr)
\!-\! \frac1{D} \Bigl(\frac{D\!-\!2}{D\!-\!1}\Bigr) R \sqrt{-g} \Biggr]
i\Delta_B(x;x') & = & i \delta^D(x-x') \; , \label{Beqn} \\
\Biggl[ \partial_{\mu} \Bigl(\sqrt{-g} g^{\mu\nu} \partial_{\nu}\Bigr)
\!-\! \frac2{D} \Bigl(\frac{D\!-\!3}{D\!-\!1}\Bigr) R \sqrt{-g} \Biggr]
i\Delta_C(x;x') & = & i \delta^D(x-x') \; . \label{Ceqn}
\end{eqnarray}
Of course the fact that we have written equations (\ref{Beqn}-\ref{Ceqn})
in an invariant form does not alter the fact that our photon propagator
only applies to a locally de Sitter background.

The various propagator equations (\ref{Aeqn}) and (\ref{Beqn}-\ref{Ceqn})
can be almost entirely solved in terms of the following function of the
invariant length $\ell(x;x')$ between $x^{\mu}$ and $x^{\prime \mu}$,
\begin{equation}
y(x;x') \equiv 4 \sin^2\Bigl(\frac12 H \ell(x;x')\Bigr) = a a' H^2
\Biggl( \Bigl\Vert \vec{x} \!-\! \vec{x}' \Bigr\Vert^2 - \Bigl(\vert\eta
\!-\! \eta'\vert - i\delta\Bigr)^2\Biggr) \; . \label{y}
\end{equation}
The most singular part of each propagator is the same as the propagator
for a massless, conformally coupled scalar \cite{BD},
\begin{equation}
{i\Delta}_{\rm cf}(x;x') = \frac{H^{D-2}}{(4\pi)^{\frac{D}2}} \Gamma\Bigl(
\frac{D}2 \!-\! 1\Bigr) \Bigl(\frac4{y}\Bigr)^{\frac{D}2-1} \; .
\end{equation}
It has long been known that no de Sitter invariant solution exists for
the $A$-type propagator \cite{AF}. It is natural to imagine our locally
de Sitter background as one of the larger class of conformally flat
geometries but with arbitrary time dependence in the scale factor. In
that case the relevant symmetry to preserve is spatial homogeneity and
isotropy, which is known as the ``E(3)'' vacuum \cite{BA}. The minimal
solution for this is \cite{OW1,OW2},
\begin{eqnarray}
\lefteqn{i \Delta_A(x;x') =  i \Delta_{\rm cf}(x;x') } \nonumber \\
& & + \frac{H^{D-2}}{(4\pi)^{\frac{D}2}} \frac{\Gamma(D \!-\! 1)}{\Gamma(
\frac{D}2)} \left\{\! \frac{D}{D\!-\! 4} \frac{\Gamma^2(\frac{D}2)}{\Gamma(D
\!-\! 1)} \Bigl(\frac4{y}\Bigr)^{\frac{D}2 -2} \!\!\!\!\!\! - \pi
\cot\Bigl(\frac{\pi}2 D\Bigr) + \ln(a a') \!\right\} \nonumber \\
& & + \frac{H^{D-2}}{(4\pi)^{\frac{D}2}} \! \sum_{n=1}^{\infty}\! \left\{\!
\frac1{n} \frac{\Gamma(n \!+\! D \!-\! 1)}{\Gamma(n \!+\! \frac{D}2)}
\Bigl(\frac{y}4 \Bigr)^n \!\!\!\! - \frac1{n \!-\! \frac{D}2 \!+\! 2}
\frac{\Gamma(n \!+\!  \frac{D}2 \!+\! 1)}{\Gamma(n \!+\! 2)} \Bigl(\frac{y}4
\Bigr)^{n - \frac{D}2 +2} \!\right\} \! . \quad \label{DeltaA}
\end{eqnarray}
On the other hand, the $B$-type and $C$-type propagators possess unique,
de Sitter invariant solutions \cite{CR,DC},
\begin{eqnarray}
i\Delta_B(x;x') & = & \frac{H^{D-2}}{(4\pi)^{\frac{D}2}} \frac{\Gamma(D\!-\!2)
\Gamma(1)}{\Gamma(\frac{D}2)} \, \mbox{}_2F_1\Bigl(D\!-\!2,1;\frac{D}2;1 \!-\!
\frac{y}4\Bigr) \; , \label{FDB} \\
i\Delta_C(x;x') & = & \frac{H^{D-2}}{(4\pi)^{\frac{D}2}} \frac{\Gamma(D\!-\!3)
\Gamma(2)}{\Gamma(\frac{D}2)} \, \mbox{}_2F_1\Bigl(D\!-\!3,2;\frac{D}2;1 \!-\!
\frac{y}4\Bigr) \; . \label{FDC}
\end{eqnarray}

One can note the progression in the coefficients of the conformal coupling
terms in the equations for $i\Delta_A$, $i\Delta_B$ and $i\Delta_C$,
\begin{equation}
-\frac{0}{D} \Bigl(\frac{D\!-\!1}{D\!-\!1}\Bigr) R \sqrt{-g} \;
\longrightarrow \;
-\frac{1}{D} \Bigl(\frac{D\!-\!2}{D\!-\!1}\Bigr) R \sqrt{-g} \;
\longrightarrow \;
-\frac{2}{D} \Bigl(\frac{D\!-\!3}{D\!-\!1}\Bigr) R \sqrt{-g} \; .
\end{equation}
The same progression is evident in the $B$-type and $C$-type solutions
(\ref{FDB}-\ref{FDC}). Were it extended to the $A$-type solution the
progression would give,
\begin{equation}
\frac{H^{D-2}}{(4\pi)^{\frac{D}2}} \frac{\Gamma(D\!-\!1) \Gamma(0)}{\Gamma(
\frac{D}2)} \, \mbox{}_2F_1\Bigl(D\!-\!1,0;\frac{D}2;1\!-\!\frac{y}4\Bigr)
\; . \label{sing}
\end{equation}
However, expression (\ref{sing}) is singular, which reflects the
incompatibility of assuming a de Sitter invariant solution.

Rather than trying to write $i\Delta_A(x;x')$ in terms of hypergeometric
functions it is actually more effective to expand $i\Delta_B(x;x')$ and
$i\Delta_C(x;x')$ in the same form as $i\Delta_A(x;x')$,
\begin{eqnarray}
\lefteqn{i \Delta_B(x;x') =  i \Delta_{\rm cf}(x;x') - \frac{H^{D-2}}{(4
\pi)^{\frac{D}2}} \! \sum_{n=0}^{\infty}\! \left\{\!  \frac{\Gamma(n \!+\! D
\!-\! 2)}{\Gamma(n \!+\! \frac{D}2)} \Bigl(\frac{y}4 \Bigr)^n \right. }
\nonumber \\
& & \hspace{6.5cm} \left. - \frac{\Gamma(n \!+\!  \frac{D}2)}{\Gamma(n \!+\!
2)} \Bigl( \frac{y}4 \Bigr)^{n - \frac{D}2 +2} \!\right\} \! , \qquad
\label{DeltaB} \\
\lefteqn{i \Delta_C(x;x') =  i \Delta_{\rm cf}(x;x') +
\frac{H^{D-2}}{(4\pi)^{\frac{D}2}} \! \sum_{n=0}^{\infty} \left\{\!
(n\!+\!1) \frac{\Gamma(n \!+\! D \!-\! 3)}{\Gamma(n \!+\! \frac{D}2)}
\Bigl(\frac{y}4 \Bigr)^n \right. } \nonumber \\
& & \hspace{4.5cm} \left. - \Bigl(n \!-\! \frac{D}2 \!+\!  3\Bigr) \frac{
\Gamma(n \!+\! \frac{D}2 \!-\! 1)}{\Gamma(n \!+\! 2)} \Bigl(\frac{y}4
\Bigr)^{n - \frac{D}2 +2} \!\right\} \! . \qquad \label{DeltaC}
\end{eqnarray}
These expressions tend to intimidate but it will be seen that they are
quite simple to use on account of the facts that the infinite sums vanish
for $D\!=\!4$ and each term goes like a positive power of $y(x;x')$. Hence
the infinite sums can only contribute when multiplied by a term which
becomes sufficiently singular upon coincidence. Note also that the $B$-type
and $C$-type propagators degenerate to $i\Delta_{\rm cf}$ in $D\!=\!4$, so
the photon propagator is the same for $D\!=\!4$ as it is in flat space!
This is the gauge that takes full advantage of the fact that electromagnetism
is conformally invariant for $D\!=\!4$.

\section{$M^2(x;x')$ in the Simplest Gauge}

The scalar self-mass-squared receives one loop contributions from a single
4-point interaction (\ref{4pt}), from a product of two 3-point interactions
(\ref{3pt}), and from the renormalization counterterms (\ref{ctm}). One
sub-section is devoted to the evaluation of each sort of contribution. By
far the most difficult is the contribution from two 3-point interactions.
That requires four separate parts!

\subsection{Contributions from the 4-Point Interaction}

The 4-point interaction (\ref{4pt}) gives rise to a diagram with the topology
depicted in Fig.~1. Its contribution is,
\begin{equation}
-i M^2_{\mbox{\tiny 4pt}}(x;x') = -i e^2 \sqrt{-g} g^{\mu\nu} i\Bigl[
\mbox{}_{\mu} \Delta_{\nu}\Bigr](x;x') \, \delta^D(x-x') \; .
\end{equation}
\begin{center}
\begin{picture}(300,80)(0,0)
\DashArrowLine(150,20)(90,20){2}
\DashArrowLine(210,20)(150,20){2}
\Vertex(150,20){3}
\Text(150,10)[b]{$x$}
\PhotonArc(150,45)(23,-90,270){3}{10.5}
\end{picture}
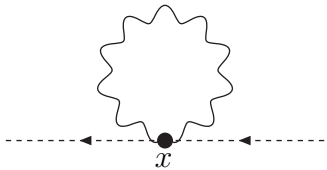
\\ {\rm Fig.~1: Contribution from the 4-point interaction.}
\end{center}
We obviously require the coincidence limits of the $B$-type and $C$-type
propagators,
\begin{eqnarray}
\lim_{x' \rightarrow x} \, {i\Delta}_B(x;x') & = & \frac{H^{D-2}}{(4\pi)^{
\frac{D}2}} \frac{\Gamma(D-1)}{\Gamma(\frac{D}2)}\times -\frac1{D\!-\!2} \; ,\\
\lim_{x' \rightarrow x} \, {i\Delta}_C(x;x') & = & \frac{H^{D-2}}{(4\pi)^{
\frac{D}2}} \frac{\Gamma(D-1)}{\Gamma(\frac{D}2)}\times \frac1{(D\!-\!2)
(D\!-\!3)} \; .
\end{eqnarray}
It follows that our result for the 4-point contribution is completely finite
in $D \!=\! 4$ dimensions,
\begin{eqnarray}
\lefteqn{-i M^2_{\mbox{\tiny 4pt}}(x;x') } \nonumber \\
& & = \frac{i e^2 H^{D-2} a^D}{(4 \pi)^{\frac{D}2}} \frac{\Gamma(D\!-\!1)}{
\Gamma(\frac{D}2)} \left\{ \Bigl(\frac{D\!  -\!1}{D\!-\!2} \Bigr) \!-\!
\frac1{(D\!-\!2)(D\!-\!3)} \right\} \delta^D(x \!-\! x') \; , \quad \\
& & \longrightarrow \frac{i e^2 H^2}{8 \pi^2} \, a^4 \delta^4(x \!-\! x') \; .
\label{4fin}
\end{eqnarray}

\subsection{Contributions from Two 3-Point Interactions}

The 3-point interaction (\ref{3pt}) gives rise to a diagram with the topology
depicted in Fig.~2. Its contribution is,
\begin{eqnarray}
-i M^2_{\mbox{\tiny 3pt}}(x;x') & = & - e^2 \sqrt{-g} g^{\mu\nu}
\sqrt{-g'} g^{\prime \rho\sigma} i\Bigl[\mbox{}_{\mu} \Delta_{\rho}\Bigr](x;x')
\partial_{\nu} \partial_{\sigma}' i\Delta_A(x;x') \nonumber \\
& & - e^2 \partial_{\nu} \Bigl[ \sqrt{-g} g^{\mu\nu} \sqrt{-g'} g^{\prime
\rho\sigma} i\Bigl[\mbox{}_{\mu} \Delta_{\rho}\Bigr](x;x') \partial_{\sigma}'
i\Delta_A(x;x') \Bigr] \nonumber \\
& & - e^2 \partial_{\sigma}' \Bigl[ \sqrt{-g} g^{\mu\nu} \sqrt{-g'} g^{\prime
\rho\sigma} i\Bigl[\mbox{}_{\mu} \Delta_{\rho}\Bigr](x;x') \partial_{\nu}
i\Delta_A(x;x') \Bigr] \nonumber \\
& & - e^2 \partial_{\nu} \partial_{\sigma}' \Bigl[ \sqrt{-g} g^{\mu\nu}
\sqrt{-g'} g^{\prime \rho\sigma} i\Bigl[\mbox{}_{\mu} \Delta_{\rho}\Bigr](x;x')
i\Delta_A(x;x') \Bigr] \; . \qquad \label{inv3}
\end{eqnarray}
\begin{center}
\begin{picture}(300,80)(0,0)
\PhotonArc(150,20)(40,0,180){5}{8}
\DashArrowLine(190,20)(110,20){2}
\DashArrowLine(110,20)(50,20){2}
\Vertex(110,20){3}
\Text(110,10)[b]{$x$}
\DashArrowLine(250,20)(190,20){2}
\Vertex(190,20){3}
\Text(192,10)[b]{$x'$}
\end{picture}
\\ {\rm Fig.~2: Contribution from two 3-point interactions.}
\end{center}

The key to an efficient calculation is that the most divergent term for each
of the three scalar propagators is the conformal propagator,
\begin{equation}
i\Delta_I(x;x') = i\Delta_{\rm cf}(x;x') + i\delta\!\Delta_I(x;x') \; .
\end{equation}
Hence the most singular part of the photon propagator has a simple tensor
structure,
\begin{equation}
i\Bigl[{}_{\mu} \Delta_{\nu}\Bigr](x;x') = a a' \eta_{\mu\nu} i\Delta_{\rm
cf}(x;x') + a a' \overline{\eta}_{\mu\nu} i\delta\!\Delta_B(x;x') - a a'
\delta^0_{\mu} \delta^0_{\nu} i\delta\!\Delta_C(x;x') \; .
\end{equation}
Now recall that only the $n\!=\!0$ term of $i\delta\!\Delta_A$ fails to vanish
in $D\!=\!4$ dimensions. Further, the $n$-th term in each $i\delta\!\Delta_I$
goes like $y^n$, so these terms can only contribute when multiplied by
something sufficiently singular from another propagator. Because this
diagram is at worst quadratically divergent --- and that only from two
factors of $i\Delta_{\rm cf}$ --- no more than one factor of
$i\delta\!\Delta_I$ can give a nonzero contribution, and only from the
$n\!=\!0$ terms. We therefore partition this subsection into computing the
respective contributions from when both the scalar and the photon propagators
provide $i\Delta_{\rm cf}$, from when the scalar provides the $n\!=\!0$ term
of $i\delta\!\Delta_A$ times $i\Delta_{\rm cf}$ from the photon propagator,
and from when the photon provides either $i\delta\!\Delta_B$ or
$i\delta\!\Delta_C$ times $i\Delta_{\rm cf}$ from the scalar propagator.

\subsubsection{Contributions from $i\Delta_{\rm cf} \times i\Delta_{\rm cf}$}
The tensor structure of the conformal part of the photon propagator gives a
simple result for the contractions in (\ref{inv3}),
\begin{eqnarray}
\lefteqn{-i M^2_{\rm cf}(x;x') = - 4e^2 (aa')^{D-1} i\Delta_{\rm cf}
\, \partial \!\cdot\! \partial^{'} i\Delta_{\rm cf} - 2e^2 a^{\prime D-1}
\partial^{\mu} \Bigl[ a^{D-1} i\Delta_{\rm cf} \Bigr] \partial_{\mu}'
i\Delta_{\rm cf} } \nonumber \\
& & \hspace{1.5cm} - 2e^2 a^{D-1} \partial^{\prime \mu} \Bigl[ a^{\prime D-1}
i\Delta_{\rm cf} \Bigr] \partial_{\mu} i\Delta_{\rm cf} - e^2 \, \partial
\!\cdot\! \partial^{\prime} \Bigl[ (aa')^{D-1} i\Delta_{\rm cf} \Bigr]
i\Delta_{\rm cf} . \qquad
\end{eqnarray}
We next substitute the explicit form of the conformal propagator,
\begin{equation}
{i\Delta}_{\rm cf}(x;x') = \frac{\Gamma\Bigl( \frac{D}2 \!-\!
1\Bigr)}{4\pi^{\frac{D}2}} \frac{(a a')^{1-\frac{D}2}}{\Delta x^{D-2}} \; .
\label{deltacf}
\end{equation}
That brings the result to form,
\begin{eqnarray}
\lefteqn{-i M^2_{\rm cf}(x;x') = - \frac{e^2 \, \Gamma^2\Bigl( \frac{D}2 \!-\!
1 \Bigr)}{16 \pi^D} \Biggl\{ \frac{4 (aa')^{\frac{D}2}}{\Delta x^{D-2}} \,
\partial \!\cdot\! \partial' \Biggl[\frac{(aa')^{1-\frac{D}2}}{\Delta x^{D-2}}
\Biggr] } \nonumber \\
& & + \frac{2 \, a'^{\frac{D}2}}{a^{\frac{D}2-1}} \, \partial^{\mu} \Biggl[
\frac{a^{\frac{D}2}}{\Delta x^{D-2}} \Biggr] \partial_{\mu}' \Biggl[
\frac{a^{\prime 1-\frac{D}2}}{\Delta x^{D-2}} \Biggr] + \frac{2 \,
a^{\frac{D}2}}{a^{\prime \frac{D}2-1}} \partial^{\prime \mu} \Biggl[
\frac{a'^{\frac{D}2}}{\Delta x^{D-2}} \Biggr] \partial_{\mu} \Biggl[
\frac{a^{1-\frac{D}2}}{\Delta x^{D-2}} \Biggr] \nonumber \\
& & \hspace{7cm} + \frac{(aa')^{1-\frac{D}2}}{\Delta x^{D-2}} \, \partial
\!\cdot\! \partial' \Biggl[\frac{(aa')^{\frac{D}2}}{\Delta x^{D-2}} \Biggr]
\Biggr\} . \qquad
\end{eqnarray}
Acting the derivatives is facilitated by the simple identities,
\begin{eqnarray}
\lefteqn{\partial_{\mu} a^p = a^p \Bigl\{\partial_{\mu} + p H a
\delta^0_{\mu}\Bigr\} \; , } \\
\lefteqn{\partial \! \cdot\! \partial' \Biggl[\frac{(aa')^p}{\Delta x^{D-2}}
\Biggr] = (aa')^{p+1} \Biggl\{\frac{(D\!-\!2) p H^2 \Delta \eta^2}{\Delta x^D}
\!-\! \frac{p^2 H^2}{\Delta x^{D-2}} \!-\! \frac{i}{a^D} \delta^D(x \!-\! x')
\Biggr\} , \qquad } \\
\lefteqn{\partial^{\mu} \Biggl[\frac{a^{p}}{\Delta x^{D-2}}\Biggr] \partial'_{
\mu} \Biggl[\frac{a^{\prime q}}{\Delta x^{D-2}}\Biggr] = a^p a^{\prime q}
\Biggl\{-\frac{(D\!-\!2)^2}{\Delta x^{2D-2}} - \frac{(D\!-\!2) q a' H
\Delta\eta}{\Delta x^{2D-2}} } \nonumber \\
& & \hspace{6.5cm} + \frac{(D\!-\! 2) p a H \Delta\eta}{\Delta x^{2D-2}}
- \frac{p q  H^2 aa'}{\Delta x^{2D-4}} \Biggr\} . \qquad
\end{eqnarray}
The result is,
\begin{eqnarray}
\lefteqn{-i M^2_{\rm cf}(x;x') = \frac{e^2 \, \Gamma^2( \frac{D}2 \!-\! 1)}{16 
\pi^{D}} \Biggl\{ \frac{4 (D\!-\!2)^2 a a'}{\Delta x^{2D-2}} } \nonumber \\
& & \hspace{2.5cm} + \frac{\frac32 (D\!-\!4)(D\!-\!2) a^2 a^{\prime 2} H^2
{\Delta \eta}^2}{\Delta x^{2D-2}} + \frac{\frac14 (D\!-\!4)^2 H^2 a^2 a^{\prime
2}}{\Delta x^{2D-4}} \Biggr\} . \qquad \label{Dres}
\end{eqnarray}

At this point it is useful to recall that the physics of $M^2(x;x')$ is
inferred by integrating it up against a smooth function in the linearized
effective field equation (\ref{effeqn}). Although each of the three terms
in (\ref{Dres}) is singular at $x^{\prime \mu} \!=\! x^{\mu}$ (and hence
$\Delta x^2 \!=\! 0$), they can each be expressed as the derivatives (with
respect to $x^{\mu}$) of functions that are integrable in $D \!=\! 4$
dimensions. Identities which facilitate this are,
\begin{eqnarray}
\frac1{\Delta x^{2D-4}} & = & \frac{\partial^2}{2(D\!-\!3)(D\!-\!4)} \Bigl(
\frac1{\Delta x^{2D-6}} \Bigr) \; , \\
\frac1{\Delta x^{2D-2}} & = & \frac{\partial^4}{4(D\!-\!2)^2 (D\!-\!3)
(D\!-\!4)} \Bigl( \frac1{\Delta x^{2D-6}} \Bigr) \; , \qquad \\
\frac{\Delta \eta^2}{\Delta x^{2D-2}} & = & \frac{\partial^2_0}{4(D\!-\!2)
(D\!-\!3)} \Bigl(\frac1{\Delta x^{2D-6}} \Bigr) \nonumber \\
& & \hspace{3cm} - \frac{\partial^2}{4(D\!-\!2) (D\!-\!3)(D\!-\!4)} 
\Bigl( \frac1{\Delta x^{2D-6}} \Bigr) \; .
\end{eqnarray}
The resulting expression for the conformal contributions to the
self-mass-squared is,
\begin{eqnarray}
\lefteqn{-i M^2_{\rm cf}(x;x') = \frac{e^2 \, \Gamma^2( \frac{D}2 \!-\!1)}{16 
\pi^{D}(D\!-\!3)} \Biggl\{ \frac{a a' \partial^4}{D\!-\!4} 
+ \frac38 (D\!-\!4) a^2 a^{\prime 2} H^2 \partial^2_0 } \nonumber \\
& & \hspace{4cm} + \frac18 (D\!-\!7) H^2 a^2 a^{\prime 2} \partial^2 \Biggr\} 
\Bigl(\frac1{\Delta x^{2D-6}} \Bigr) . \qquad \label{pint}
\end{eqnarray}

All of the expressions in (\ref{pint}) are integrable with respect to 
$x^{\prime \mu}$ in the effective field equation (\ref{effeqn}). We could 
take $D\!=\!4$ at this stage were it not for the explicit factor of 
$1/(D\!-\!4)$ in the first term. This is an ultraviolet divergence, however, 
it is not yet in the correct form to be removed by a local counterterm. We 
can achieve this form by employing the identity,
\begin{equation}
\partial^2 \Bigl(\frac1{\Delta x^{D-2}} \Bigr) = \frac{i 4 \pi^{\frac{D}2}}{
\Gamma(\frac{D}2 \!-\! 1)} \delta^D(x\!-\!x') \; .
\end{equation}
The procedure is to first add zero to the basic divergence,
\begin{equation}
\frac{\partial^2}{D\!-\!4} \Bigl(\frac1{\Delta x^{2D-6}} \Bigr) = 
\frac{\partial^2}{D\!-\!4} \Bigl(\frac1{\Delta x^{2D-6}} - \frac{\mu^{D-4}}{
\Delta x^{D-2}} \Bigr) + \frac{\mu^{D-4}}{D\!-\!4} \times 
\frac{i 4 \pi^{\frac{D}2}}{\Gamma(\frac{D}2 \!-\! 1)} \delta^D(x\!-\!x') \; .
\end{equation}
The parenthesized nonlocal term is not only integrable in $D\!=\!4$, it
also vanishes in that dimension,
\begin{eqnarray}
\frac1{\Delta x^{2D-6}} - \frac{\mu^{D-4}}{\Delta x^{D-2}} & = & 
\frac{\mu^{2D-8}}{\Delta x^2} \left\{ \Bigl(\frac1{\mu^2 \Delta x^2}
\Bigr)^{D-4} - \Bigl(\frac1{\mu^2 \Delta x^2}\Bigr)^{\frac{D}2-2} \right\} 
\; , \\
& = & -\frac12 (D\!-\!4) \, \frac{\ln(\mu^2 \Delta x^2)}{\Delta x^2} +
O\Bigl( (D\!-\!4)^2\Bigr) \; . 
\end{eqnarray}
We can therefore segregate the basic divergence on a local term --- which can
be absorbed into a counterterm --- and take the limit $D\!=\!4$ on the
finite, nonlocal term,
\begin{equation}
\frac{\partial^2}{D\!-\!4} \Bigl(\frac1{\Delta x^{2D-6}} \Bigr) = 
\frac{i 4 \pi^{\frac{D}2} \mu^{D-4}}{\Gamma(\frac{D}2 \!-\!1)} \frac{
\delta^D(x\!-\!x')}{D\!-\!4} -\frac12 \partial^2 \Bigl( 
\frac{\ln(\mu^2 \Delta x^2)}{\Delta x^2} \Bigr) + O(D\!-\!4) \; . \qquad
\end{equation}
Employing this in (\ref{pint}) and taking the other terms to $D\!=\!4$ gives,
\begin{eqnarray}
\lefteqn{-i M^2_{\rm cf}(x;x') = \frac{i e^2 \mu^{D-4}}{4 \pi^{\frac{D}2}} 
\frac{\Gamma( \frac{D}2 \!-\! 1)}{(D\!-\!3) (D\!-\!4)} \, a a' \partial^2
\delta^D(x\!-\!x') } \nonumber \\
& & \hspace{1cm} - \frac{i 3 e^2 H^2}{2^5 \pi^2} \, a^4 \delta^4(x\!-\!x')
- \frac{e^2}{2^5 \pi^4} \, a a' \partial^4 \Bigl( \frac{\ln(\mu^2 \Delta x^2)}{
\Delta x^2} \Bigr) + O(D\!-\!4) \; . \qquad \label{cffin}
\end{eqnarray}

\subsubsection{Contributions from $i\Delta_{\rm cf} \times i\delta\!\Delta_A$}

Recall that the full scalar propagator is $i\Delta_A(x;x') = 
i\Delta_{\rm cf}(x;x') + i\delta\!\Delta_A(x;x')$. We have just computed the
result of keeping only the conformal parts of the photon and scalar
propagators. Using just the conformal part of the photon propagator and 
setting the scalar propagator to $i\delta\!\Delta_A$ gives,
\begin{eqnarray}
\lefteqn{-i M^2_{\delta A}(x;x') = - e^2 (aa')^{D-1} i\Delta_{\rm cf} \,
\partial \!\cdot\! \partial' i\delta\!\Delta_A - e^2 a^{\prime D-1} 
\partial^{\mu} \Bigl[a^{D-1} i\Delta_{\rm cf} \, \partial_{\mu}' 
i\delta\!\Delta_A \Bigr] } \nonumber \\
& & \hspace{1.3cm} - e^2 a^{D-1} \partial^{\prime \mu} \Bigl[ a^{\prime D-1} 
i\Delta_{\rm cf} \, \partial_{\mu} i\delta\!\Delta_A \Bigr] - e^2 \partial 
\!\cdot\! \partial' \Bigl[ (aa')^{D-1} i\Delta_{\rm cf} \, i\delta\!\Delta_A 
\Bigr] . \qquad \label{dAcon}
\end{eqnarray}
It is useful to separately consider the reduction of each of the four terms 
on the right-hand side of (\ref{dAcon}), labeling them ``1'', ``2'', ``3'' 
and ``4'', respectively.

We can read off $i\delta\!\Delta_A$ from the $A$-type propagator 
(\ref{DeltaA}),
\begin{eqnarray}
\lefteqn{i\delta\!\Delta_A(x;x') = } \nonumber \\
& & \frac{H^2}{16 \pi^{\frac{D}2}} \frac{\Gamma(\frac{D}2 \!+\! 1)}{\frac{D}2
\!-\! 2} \frac{(a a')^{2- \frac{D}2}}{\Delta x^{D-4}} 
+ \frac{H^{D-2}}{(4\pi)^\frac{D}2} \frac{\Gamma(D\!-\!1)}{\Gamma( \frac{D}2
)} \Biggl\{- \pi\cot\Bigl(\frac{\pi}2 D\Bigr) + \ln(aa') \Biggr\} \nonumber \\
& & + \frac{H^{D-2}}{(4\pi)^{\frac{D}2}} \! \sum_{n=1}^{\infty}\! \left\{\!
\frac1{n} \frac{\Gamma(n \!+\! D \!-\! 1)}{\Gamma(n \!+\! \frac{D}2)}
\Bigl(\frac{y}4 \Bigr)^n \!\!\!\! - \frac1{n \!-\! \frac{D}2 \!+\! 2}
\frac{\Gamma(n \!+\!  \frac{D}2 \!+\! 1)}{\Gamma(n \!+\! 2)} \Bigl(\frac{y}4
\Bigr)^{n - \frac{D}2 +2} \!\right\} \! . \quad \label{dA}
\end{eqnarray}
In $D\!=\!4$ the most singular contributions to (\ref{dAcon}) have the
form, $i\delta\!\Delta_A/{\Delta x}^4$. Because the infinite series 
terms in (\ref{dA}) go like positive powers of $\Delta x^2$ these terms make 
integrable contributions to (\ref{dAcon}). We can therefore take $D\!=\!4$, 
at which point we see that all the infinite series terms drop. It is sometimes 
also useful to note that the $D=4$ limit is,
\begin{equation}
i\delta\!\Delta_A(x;x') = -\frac{H^2}{8 \pi^2} \ln\Bigl(\frac{\sqrt{e}}4 H^2
\Delta x^2\Bigr) + O(D\!-\!4) \; .
\end{equation}
Note that the factor of $e$ in this expression is the base of the natural
logarithm rather than the electromagnetic coupling constant! The distinction
should always be clear from context.

We begin the reduction of the first term in (\ref{dAcon}) by substituting 
the conformal propagator,
\begin{eqnarray}
-i M^2_{\delta A_1}(x;x') & \equiv & - e^2 (aa')^{D-1} i\Delta_{\rm cf}(x;x')
\, \partial \!\cdot\! \partial' i\delta\!\Delta_A(x;x') \; , \\
& = & -\frac{e^2 \Gamma(\frac{D}2 \!-\!1)}{4 \pi^{\frac{D}2}} 
\frac{(a a')^{\frac{D}2}}{\Delta x^{D-4}}
\, \partial \!\cdot\! \partial' i\delta\!\Delta_A(x;x') \; .
\end{eqnarray}
Only the first term of (\ref{dA}) survives when acted upon by
$\partial \!\cdot\! \partial'$,
\newpage
\begin{eqnarray}
\lefteqn{-i M^2_{\delta A_1}(x;x') } \nonumber \\
& & = -\frac{e^2 H^2}{2^6 \pi^D} \frac{\Gamma(\frac{D}2 \!-\! 1) 
\Gamma(\frac{D}2 \!+\! 1)}{\frac{D}2 \!-\! 2} \frac{(a a')^{\frac{D}2}}{
\Delta x^{D-2}} \, \partial \!\cdot\! \partial' \Biggl[ \frac{(a a')^{2-
\frac{D}2}}{\Delta x^{D-4}} \Biggr] + O(D\!-\!4)\; , \\
& & = \frac{e^2 H^2}{2^6 \pi^D} \Gamma\Bigl(\frac{D}2 \!-\! 1\Bigr)
\Gamma\Bigl(\frac{D}2 \!+\! 1\Bigr) (a a')^2 \Biggl\{-\frac4{\Delta x^{2D-4}}
+ \frac{(D\!-\!4) a a' H^2 \Delta \eta^2}{\Delta x^{2D-4}} \nonumber \\
& & \hspace{6.5cm} + \frac{(D\!-\!4) a a' H^2}{2 \Delta x^{2D-6}} \Biggr\} 
+ O(D\!-\!4) \; . \qquad
\end{eqnarray}
The first term in this expression is of a type we have considered in the
previous sub-section. On the other hand, the last two terms are integrable,
so they give zero contribution on account of the factors of $D\!-\!4$. The
final result is,
\begin{eqnarray}
\lefteqn{-i M^2_{\delta A_1}(x;x') = -\frac{i e^2 H^2 \mu^{D-4}}{8
\pi^{\frac{D}2}} \frac{\Gamma(\frac{D}2 \!+\! 1)}{(D\!-\!3) (D\!-\!4)} 
\, a^4 \delta^D(x\!-\!x') } \nonumber \\
& & \hspace{4.5cm} + \frac{e^2 H^2}{2^5 \pi^4} (a a')^2 \partial^2 \Bigl( 
\frac{\ln(\mu^2 \Delta x^2)}{\Delta x^2} \Bigr) + O(D\!-\!4) \; . \qquad
\label{dA1}
\end{eqnarray}

The second term in (\ref{dAcon}) is,
\begin{eqnarray}
\lefteqn{-i M^2_{\delta A_2}(x;x') \equiv - e^2 a^{\prime D-1} \partial^{\mu} 
\Bigl[a^{D-1} i\Delta_{\rm cf}(x;x') \, \partial_{\mu}' i\delta\!\Delta_A(x;x')
\Bigr] \; , } \\
& & \hspace{1cm} = -\frac{e^2 \Gamma(\frac{D}2 \!-\!1)}{4 \pi^{\frac{D}2}} 
(a a')^{\frac{D}2} \Biggl\{ \partial^{\mu} \Biggl[ \frac{\partial_{\mu}'
i\delta\!\Delta_A}{\Delta x^{D-2}} \Biggr] - \frac{D H a}{2 \Delta x^{D-2}} 
\, \partial_0' i\delta\!\Delta_A \Biggr\} . \qquad
\end{eqnarray}
The derivative of $i\delta\!\Delta_A(x;x')$ is,
\begin{eqnarray}
\lefteqn{\partial_{\mu}' i\delta\!\Delta_A(x;x') = 
\frac{H^2}{16 \pi^{\frac{D}2}} \Gamma\Bigl(\frac{D}2 \!+\! 1\Bigr) 
(a a')^{2-\frac{D}2} \Biggl[ \frac{2 \Delta x_{\mu}}{\Delta x^{D-2}} -
\frac{H a' \delta^0_{\mu}}{\Delta x^{D-4}} \Biggr] } \nonumber \\
& & \hspace{5cm} + \frac{H^{D-2}}{2^D \pi^{\frac{D}2}} \frac{\Gamma(D\!-\!1)}{
\Gamma(\frac{D}2)} \, H a' \delta_{\mu}^0 + O(D\!-\!4) \; . \qquad
\end{eqnarray}
Combining the two results gives,
\begin{eqnarray}
\lefteqn{-i M^2_{\delta A_2}(x;x') =  \frac{e^2 H^2}{2^6 \pi^D} \Gamma\Bigl(
\frac{D}2 \!-\! 1\Bigr) \Gamma\Bigl(\frac{D}2 \!+\! 1\Bigr) } \nonumber \\
& & \hspace{1cm} \times (a a')^2 \Biggl\{\frac{2(D\!-\!4)}{\Delta x^{2D-4}} 
\!-\! \frac{2(D\!-\!3) a' H \Delta \eta}{\Delta x^{2D-4}} \!-\! \frac{4 a H 
\Delta \eta}{\Delta x^{2D-4}} \!-\! \frac{2 a a' H^2}{\Delta x^{2D-6}} 
\Biggr\} \nonumber \\
& & + \frac{e^2 H^{D-2}}{2^{D+2} \pi^D} \, \Gamma(D \!-\! 1) 
(a a')^{\frac{D}2} \Biggl\{ \frac{2 a' H \Delta \eta}{\Delta x^D} \!+\!
\frac{D a a' H^2}{(D\!-\!2) \Delta x^{D-2}} \Biggr\} + O(D\!-\!4) \; . \qquad
\label{dA2}
\end{eqnarray}

Of course the third term in (\ref{dAcon}) follows from the second just by
interchanging $x^{\mu}$ and $x^{\prime \mu}$,
\begin{eqnarray}
\lefteqn{-i M^2_{\delta A_3}(x;x') =  \frac{e^2 H^2}{2^6 \pi^D} \Gamma\Bigl(
\frac{D}2 \!-\! 1\Bigr) \Gamma\Bigl(\frac{D}2 \!+\! 1\Bigr) } \nonumber \\
& & \hspace{1cm} \times (a a')^2 \Biggl\{\frac{2(D\!-\!4)}{\Delta x^{2D-4}} 
\!+\! \frac{2(D\!-\!3) a H \Delta \eta}{\Delta x^{2D-4}} \!+\! \frac{4 a' H
\Delta \eta}{\Delta x^{2D-4}} \!-\! \frac{2 a a' H^2}{\Delta x^{2D-6}} 
\Biggr\} \nonumber \\
& & + \frac{e^2 H^{D-2}}{2^{D+2} \pi^D} \, \Gamma(D \!-\! 1) 
(a a')^{\frac{D}2} \Biggl\{\!- \frac{2 a H \Delta \eta}{\Delta x^D} \!+\!
\frac{D a a' H^2}{(D\!-\!2) \Delta x^{D-2}} \!\Biggr\} + O(D\!-\!4) \; . \qquad
\label{dA3}
\end{eqnarray}
The fact that $a \!-\! a' = a a' H {\Delta \eta}$ allows us to usefully combine
(\ref{dA2}) and (\ref{dA3}),
\begin{eqnarray}
\lefteqn{-i M^2_{\delta A_{2+3}}(x;x') = \frac{e^2 H^2}{2^6 \pi^D} \Gamma\Bigl(
\frac{D}2 \!-\! 1\Bigr) \Gamma\Bigl(\frac{D}2 \!+\! 1\Bigr) } \nonumber \\
& & \hspace{1cm} \times (a a')^2 \Biggl\{\frac{4(D\!-\!4)}{\Delta x^{2D-4}} 
\!+\! \frac{2(D\!-\!5) a a' H^2 \Delta \eta^2}{\Delta x^{2D-4}} \!-\! 
\frac{4 a a' H^2}{\Delta x^{2D-6}} \Biggr\} \nonumber \\
& & \hspace{-.5cm} + \frac{e^2 H^{D-2}}{2^{D+2} \pi^D} \, \Gamma(D \!-\! 1) 
(a a')^{\frac{D}2+1} \Biggl\{\!- \frac{2 H^2 \Delta \eta^2}{\Delta x^D} \!+\!
\frac{2 D H^2}{(D\!-\!2) \Delta x^{D-2}} \!\Biggr\} + O(D\!-\!4) \; . \qquad 
\end{eqnarray}
Only the first term is not immediately integrable, and the explicit factor
of $D\!-\!4$ it bears results in a finite, local contribution from it,
\begin{equation}
\frac{4(D\!-\!4)}{\Delta x^{2D-4}} = \frac{2 \partial^2}{D\!-\!3} \Bigl(
\frac1{\Delta x^{2D-6}} \Bigr) = i 8 \pi^2 \delta^4(x\!-\!x') + O(D\!-\!4) \; .
\end{equation} 
Hence the second and third terms in (\ref{dAcon}) make a completely finite
contribution,
\begin{equation}
-i M^2_{\delta A_{2+3}}\!(x;x') \!=\! \frac{i e^2 H^2}{8 \pi^2} a^4 
\delta^4(x\!-\!x') + \frac{e^2 H^4}{2^6 \pi^4} \, (a a')^3 (2 \partial_0^2
+ \partial^2) \ln(\Delta x^2) + O(D\!-\!4) . \label{dA2+3}
\end{equation}

The final term in (\ref{dAcon}) has the simplest reduction and the most
interesting eventual contribution. Its reduction is accomplished by simply
refraining from acting the external derivatives and instead taking the
limit $D\!=\!4$ on the integrable expression inside the square brackets,
\begin{eqnarray}
\lefteqn{-i M^2_{\delta A_4}(x;x') \equiv - e^2 \partial \!\cdot\! \partial'
\Bigl[(a a')^{D-1} i\Delta_{\rm cf}(x;x') \, i\delta\!\Delta_A(x;x')
\Bigr] \; , } \\
& & = \frac{e^2 H^2}{2^5 \pi^4} \, \partial \!\cdot\! \partial' \Biggl\{
\frac{(a a')^2}{\Delta x^2} \, \ln\Bigl(\frac{\sqrt{e}}4 H^2 \Delta x^2\Bigr)
\Biggr\} + O(D\!-\!4) \; , \\
& & = -\frac{e^2 H^2}{2^5 \pi^4} \, (a a')^2 \partial^2 \Biggl\{
\frac{\ln\Bigl(\frac{\sqrt{e}}4 H^2 \Delta x^2\Bigr)}{\Delta x^2} \Biggr\} 
- \frac{e^2 H^4}{2^6 \pi^4} \, (a a')^3 \Biggl\{ \partial_0^2
\Biggl[ \ln^2\Bigl(\frac{\sqrt{e}}4 H^2 \Delta x^2\Bigr) \Biggr] \nonumber \\
& & \hspace{2cm} + \frac32 \partial^2 \Biggl[ \ln^2\Bigl(\frac{\sqrt{e}}4 H^2 
\Delta x^2\Bigr) \!-\! 2 \ln\Bigl(\frac{\sqrt{e}}4 H^2 \Delta x^2\Bigr) \Biggr]
\Biggr\} + O(D\!-\!4) \; . \quad \label{dA4}
\end{eqnarray}

We can now sum (\ref{dA1}), (\ref{dA2+3}) and (\ref{dA4}) to obtain the
final result for this sub-section,
\begin{eqnarray}
\lefteqn{-i M^2_{\delta A}(x;x') = \frac{i e^2 H^2 \mu^{D-4}}{2^4 
\pi^{\frac{D}2}} \Biggl\{- \frac{2\Gamma(\frac{D}2\!+\!1)}{(D\!-\!3) (D\!-\!4)}
- 4 \ln\Bigl(\frac{H}{2 \mu}\Bigr) + 1 \Biggr\} a^4 \delta^D(x\!-\!x') }
\nonumber \\
& & \hspace{1.5cm} - \frac{e^2 H^4}{2^6 \pi^4} \, (a a')^3 \Biggl\{ 
\partial_0^2 \Bigl[ \ln^2\Bigl(2^{-2} H^2 \Delta x^2\Bigr) \!-\!
\ln\Bigl(2^{-2} H^2 \Delta x^2\Bigr) \Bigr] \nonumber \\
& & \hspace{1.7cm} + \frac{\partial^2}2 \Bigl[3 \ln^2\Bigl(2^{-2}
H^2\Delta x^2\Bigr) \!-\! 5 \ln\Bigl(2^{-2} H^2 \Delta x^2\Bigr) \Bigr] 
\Biggr\} + O(D\!-\!4) \; .\qquad \label{dAfin}
\end{eqnarray}
Note that all the nonlocal terms proportional to $(a a')^2$ have combined to
produce a finite, local term.

\subsubsection{Contributions from $i\delta\!\Delta_B \times i\Delta_{\rm cf}$}

In this sub-section we evaluate the contributions to the full 3-point
diagram (\ref{inv3}) by making the following replacements for the scalar
and photon propagators,
\begin{eqnarray}
i\Delta_A(x;x') & \longrightarrow & i\Delta_{\rm cf}(x;x') \; , \label{rep1} \\
i\Bigl[\mbox{}_{\mu}\Delta_{\rho}\Bigr](x;x') & \longrightarrow & a a' \,
i\delta\!\Delta_B(x;x') \, \overline{\eta}_{\mu\rho} \; . \label{rep2}
\end{eqnarray}
Here $\overline{\eta}_{\mu\rho}$ is the purely spatial part (\ref{bar}) of
the Lorentz metric and $i\delta\!\Delta_B(x;x')$ is the residual of the
$B$-type propagator (\ref{DeltaB}) after the conformal contribution has
been subtracted,
\begin{eqnarray}
\lefteqn{i\delta\!\Delta_B(x;x') = \frac{H^2 \Gamma(\frac{D}2)}{16 
\pi^{\frac{D}2}} \frac{(a a')^{2-\frac{D}2}}{\Delta x^{D-4}} 
-\frac{H^{D-2}}{(4\pi)^\frac{D}2} \frac{\Gamma(D\!-\!2)}{\Gamma\Bigl(\frac{D}2 
\Bigr)} } \nonumber \\
& & \hspace{2cm} + \frac{H^{D-2}}{(4 \pi)^{\frac{D}2}} \sum_{n=1}^{\infty} 
\left\{ \frac{\Gamma(n \!+\!  \frac{D}2)}{\Gamma(n \!+\! 2)} \Bigl( \frac{y}4 
\Bigr)^{n - \frac{D}2 +2} - \frac{\Gamma(n \!+\! D \!-\! 2)}{\Gamma(n \!+\! 
\frac{D}2)} \Bigl(\frac{y}4 \Bigr)^n \right\} \; . \qquad \label{dB}
\end{eqnarray}
As was the case for the $i\delta\!\Delta_A(x;x')$ contributions considered 
in the previous sub-section, this diagram is not sufficiently singular for
the infinite series terms from $i\delta\!\Delta_B(x;x')$ to make a nonzero
contribution in the $D\!=\!4$ limit. Unlike $i\delta\!\Delta_A(x;x')$, even
the $n\!=\!0$ terms of $i\delta\!\Delta_B(x;x')$ vanish for $D\!=\!4$. This
means they can only contribute when multiplied by a divergence.

Making replacements (\ref{rep1}) and (\ref{rep2}) in the 3-point diagram 
(\ref{inv3}) gives,
\begin{eqnarray}
\lefteqn{-i M^2_{\delta B}(x;x') = - e^2 (aa')^{D-1} \Biggl\{i\delta\!\Delta_B 
\partial_i \partial^{'}_i i\Delta_{\rm cf} + \partial_i \Bigl[ 
i\delta\!\Delta_B \partial_i' i\delta\Delta_{\rm cf} \Bigr] } \nonumber \\
& & \hspace{4.5cm} + \partial_i' \Bigl[ i\delta\Delta_B \partial_i 
i\Delta_{\rm cf} \Bigr] + \partial_i \partial_i' \Bigl[ i\delta\Delta_B 
i\Delta_{\rm cf}\Bigr] \Biggr\} . \qquad
\end{eqnarray}
Because all the derivatives are spatial we can extract the scale factors 
from $i\Delta_{\rm cf}(x;x')$. We can also convert $\partial_i'$ to 
$-\partial_i$ and combine two terms to obtain,
\begin{eqnarray}
\lefteqn{-i M^2_{\delta B}(x;x') = \frac{e^2 \Gamma(\frac{D}2\!-\! 1)}{4 
\pi^{\frac{D}2}} \, (a a')^{\frac{D}2} } \nonumber \\
& & \hspace{.7cm} \times \Biggl\{i\delta\!\Delta_B \nabla^2 \Bigl( 
\frac1{\Delta x^{D-2}} \Bigr) + 2 \partial_i \Bigl[ i\delta\!\Delta_B 
\partial_i \Bigl( \frac1{\Delta x^{D-2}} \Bigr) \Bigr] + \nabla^2 \Bigl( 
\frac{i\delta\!\Delta_B}{\Delta x^{D-2}} \Bigr) \Biggr\} . \qquad \label{dBcon}
\end{eqnarray}
In analogy with the analysis of the previous sub-section we can label
the three terms of this expression ``1'', ``$2\!+\!3$'' and ``4''.

The only nonzero contribution comes from the first term of (\ref{dBcon}). To 
get it one merely substitutes the first line of (\ref{dB}) and then exploits 
some simple derivative identities,
\begin{eqnarray}
\lefteqn{-i M^2_{\delta B_1}(x;x') = \frac{e^2 H^2}{2^6 \pi^D} \Gamma\Bigl(
\frac{D}2\!-\!1\Bigr) \Gamma\Bigl(\frac{D}2\Bigr) \frac{(a a')^2}{\Delta 
x^{D-4}} \nabla^2 \Bigl(\frac1{\Delta x^{D-2}} \Bigr) } \nonumber \\
& & \hspace{2.5cm} -\frac{e^2 H^{D-2}}{2^{D+2} \pi^D} \frac{\Gamma(D\!-\!2)}{
\frac{D}2\!-\!1} \, (a a')^{\frac{D}2} \nabla^2 \Bigl( \frac1{\Delta x^{D-2}} 
\Bigr) + O(D\!-\!4) \; , \qquad \\
& & = \frac{e^2 H^2}{2^6 \pi^D} \frac{\Gamma(\frac{D}2\!-\!1) 
\Gamma(\frac{D}2)}{4 (D\!-\!3)} \, (a a')^2 \Bigl[ D\nabla^2 \!-\! (D\!-\!1) 
\partial^2 \Bigr] \Bigl( \frac1{\Delta x^{2D-6}} \Bigr) \nonumber \\
& & \hspace{2.5cm} -\frac{e^2 H^{D-2}}{2^{D+2} \pi^{D}} \frac{\Gamma(D\!-\!
2)}{\frac{D}2 \!-\!1} \, (a a')^{\frac{D}2} \nabla^2 \Bigl( \frac1{\Delta 
x^{D-2}} \Bigr) + O(D\!-\!4) \; .  \qquad
\end{eqnarray}
The last expression is completely integrable so we can take the limit 
$D\!=\!4$, at which stage the spatial derivative terms cancel and the 
d'Alembertian term gives a delta function,
\begin{equation}
-i M^2_{\delta B_1}(x;x') = 
-\frac{i 3 e^2 H^2}{2^6 \pi^2} \, a^4 \delta^4(x \!-\! x') + O(D\!-\!4) \; .
\end{equation}

The middle term of (\ref{dBcon}) is integrable with $\partial_i$ extracted,
so we can take $D\!=\!4$ immediately. This gives zero because 
$i\delta\!\Delta_B$ vanishes in $D\!=\!4$,
\begin{equation}
-i M^2_{\delta B_{2+3}}(x;x') \equiv \frac{e^2 \Gamma(\frac{D}2\!-\! 1)}{4 
\pi^{\frac{D}2}} \, (a a')^{\frac{D}2} \, 2 \partial_i \Bigl[ i\delta\!\Delta_B 
\partial_i \Bigl( \frac1{\Delta x^{D-2}} \Bigr) \Bigr] = O(D\!-\!4) \; .
\end{equation}
The same analysis and the same result pertains for the final term of 
(\ref{dBcon}),
\begin{equation}
-i M^2_{\delta B_4}(x;x') \equiv \frac{e^2 \Gamma(\frac{D}2\!-\! 1)}{4 
\pi^{\frac{D}2}} \, (a a')^{\frac{D}2} \, \nabla^2 \Bigl( \frac{i\delta\!
\Delta_B}{\Delta x^{D-2}} \Bigr) = O(D\!-\!4) \; .
\end{equation}
Our final result for this sub-section is therefore,
\begin{equation}
-i M^2_{\delta B}(x;x') = -\frac{i 3 e^2 H^2}{2^6 \pi^2} \, a^4 \delta^4(x 
\!-\! x') + O(D\!-\!4) \; . \label{dBfin}
\end{equation}

\subsubsection{Contributions from $i\delta\!\Delta_C \times i\Delta_{\rm cf}$}

In this sub-section we evaluate the contributions to the full 3-point
diagram (\ref{inv3}) by making the following replacements for the scalar
and photon propagators,
\begin{eqnarray}
i\Delta_A(x;x') & \longrightarrow & i\Delta_{\rm cf}(x;x') \; ,\label{Crep1} \\
i\Bigl[\mbox{}_{\mu}\Delta_{\rho}\Bigr](x;x') & \longrightarrow & -a a' \,
i\delta\!\Delta_C(x;x') \, \delta^0_{\mu} \delta^0_{\rho} \; . \label{Crep2}
\end{eqnarray}
The result is,
\begin{eqnarray}
\lefteqn{-i M^2_{\delta C}(x;x') = e^2 (aa')^{D-1} i\delta\!\Delta_C 
\partial_0 \partial_0' i\Delta_{\rm cf} + e^2 a^{\prime D-1} \partial_0 
\Bigl[ a^{D-1} i\delta\!\Delta_C \partial_0' i\Delta_{\rm cf} \Bigr] }
\nonumber \\
& & \hspace{1.5cm} + e^2 a^{D-1} \partial_0' \Bigl[ a^{\prime D-1} 
i\delta\!\Delta_C \partial_0 i\Delta_{\rm cf} \Bigr] + e^2 \partial_0 
\partial_0' \Bigl[ (aa')^{D-1} i\delta\!\Delta_C i\Delta_{\rm cf} \Bigr] . 
\qquad \label{dCcon}
\end{eqnarray}
Here $i\delta\!\Delta_C(x;x')$ is the residual of the $C$-type propagator 
(\ref{DeltaC}) after the conformal contribution has been subtracted,
\begin{eqnarray}
\lefteqn{i \delta\!\Delta_C(x;x') = \frac{H^2}{16 \pi^{\frac{D}2}} 
\Bigl( \frac{D}2 \!-\! 3\Bigr) \Gamma\Bigl(\frac{D}2 \!-\! 1\Bigr) 
\frac{(a a')^{2-\frac{D}2}}{\Delta x^{D-4}}+ \frac{H^{D-2}}{(4\pi)^{\frac{D}2}}
\frac{\Gamma(D \!-\! 3)}{\Gamma(\frac{D}2)} } \nonumber \\
& & \hspace{-.7cm} - \frac{H^{D-2}}{(4\pi)^{\frac{D}2}} \!\!\sum_{n=1}^{\infty} 
\!\!\left\{ \!\!\Bigl(n \!-\! \frac{D}2 \!+\! 3\Bigr) \frac{\Gamma(n \!+\!
\frac{D}2 \!-\! 1)}{\Gamma(n \!+\! 2)} \Bigl(\frac{y}4 \Bigr)^{n -\frac{D}2 +2}
\!\!\!\!\!\!\! - (n\!+\!1) \frac{\Gamma(n \!+\! D \!-\! 3)}{\Gamma(n \!+\! 
\frac{D}2)} \Bigl(\frac{y}4 \Bigr)^n \!\right\} \!. \qquad \label{dC}
\end{eqnarray}

As with the contributions from $i\delta\!\Delta_B(x;x')$ considered in the
previous sub-section, the only way $i\delta\!\Delta_C(x;x')$ can give a
nonzero contribution in $D\!=\!4$ dimensions is for it to multiply a singular
term. For (\ref{dCcon}) that means only the $n\!=\!0$ term can possibly 
contribute. Even for the $n\!=\!0$ term, both derivatives must act upon a
$\Delta x^2$. If even a single derivative acts instead upon a scale factor, 
the result is a term which is integrable in $D\!=\!4$ dimensions, at which 
point the cancellations between pairs of terms evident in (\ref{dC}) results 
in zero net contribution. We can therefore extract the scale factors from 
$i\Delta_{\rm cf}(x;x')$ and replace $\partial_0'$ by $-\partial_0$ as we did 
in the previous sub-section,
\begin{eqnarray}
\lefteqn{-i M^2_{\delta C}(x;x') = -\frac{e^2 \Gamma(\frac{D}2\!-\! 1)}{4 
\pi^{\frac{D}2}} \, (a a')^{\frac{D}2} \Biggl\{i\delta\!\Delta_C \partial_0^2 
\Bigl( \frac1{\Delta x^{D-2}} \Bigr) } \nonumber \\
& & \hspace{2cm} + 2 \partial_0 \Bigl[ i\delta\!\Delta_C \partial_0 \Bigl( 
\frac1{\Delta x^{D-2}} \Bigr) \Bigr] + \partial_0^2 \Bigl( \frac{i\delta\!
\Delta_C}{\Delta x^{D-2}} \Bigr) \Biggr\} + O(D\!-\!4) \; . \qquad 
\label{dCCcon}
\end{eqnarray}

The first term in (\ref{dCcon}) is reduced the same way as was the analogous
first term from $i\delta\!\Delta_B(x;x')$ in the previous sub-section,
\begin{eqnarray}
\lefteqn{-i M^2_{\delta C_1}(x;x') \equiv -\frac{e^2 \Gamma(\frac{D}2\!-\!1)}{4 
\pi^{\frac{D}2}} \, (a a')^{\frac{D}2} i\delta\!\Delta_C \partial_0^2 
\Bigl( \frac1{\Delta x^{D-2}} \Bigr) \; ,} \\
& & = -\frac{e^2 H^2}{2^6 \pi^D} \Bigl(\frac{D}2\!-\!3\Bigr) 
\Gamma^2\Bigl(\frac{D}2 \!-\! 1\Bigr) \frac{(a a')^2}{\Delta x^{D-4}} 
\partial_0^2 \Bigl(\frac1{\Delta x^{D-2}} \Bigr) \nonumber \\
& & \hspace{2.5cm} -\frac{e^2 H^{D-2}}{2^{D+2} \pi^D} \frac{\Gamma(D\!-\!3)}{
\frac{D}2\!-\!1} \, (a a')^{\frac{D}2} \partial_0^2 \Bigl( 
\frac1{\Delta x^{D-2}} \Bigr) + O(D\!-\!4) \; , \qquad \\
& & = -\frac{e^2 H^2}{2^6 \pi^D} \frac{(\frac{D}2\!-\!3) \Gamma^2(\frac{D}2
\!-\!1)}{4 (D\!-\!3)} \, (a a')^2 \Bigl[ D \partial_0^2 \!+\! \partial^2 \Bigr]
\Bigl( \frac1{\Delta x^{2D-6}} \Bigr) \nonumber \\
& & \hspace{2.5cm} -\frac{e^2 H^{D-2}}{2^{D+2} \pi^{D}} \frac{\Gamma(D\!-\!
3)}{\frac{D}2 \!-\!1} \, (a a')^{\frac{D}2} \partial_0^2 \Bigl( \frac1{\Delta 
x^{D-2}} \Bigr) + O(D\!-\!4) \; , \\
& & = \frac{e^2 H^2}{2^8 \pi^4} \, (a a')^2 \partial^2 
\Bigl( \frac1{\Delta x^2} \Bigr) + O(D\!-\!4) \; , \\
& & = \frac{i e^2 H^2}{2^6 \pi^2} \, a^4 \delta^4(x\!-\!x') + O(D\!-\!4) \; .
\end{eqnarray}
As in the previous sub-section, the other terms are zero because they start
out with enough derivatives extracted that we can take $D\!=\!4$ right away,
\begin{equation}
-i M^2_{\delta C_{2+3}}(x;x') \equiv -\frac{e^2 \Gamma(\frac{D}2\!-\! 1)}{4 
\pi^{\frac{D}2}} \, (a a')^{\frac{D}2} \, 2 \partial_0 \Bigl[ i\delta\!\Delta_C
\partial_0 \Bigl( \frac1{\Delta x^{D-2}} \Bigr) \Bigr] = O(D\!-\!4) \; ,
\end{equation}
\begin{equation}
-i M^2_{\delta C_4}(x;x') \equiv -\frac{e^2 \Gamma(\frac{D}2\!-\! 1)}{4 
\pi^{\frac{D}2}} \, (a a')^{\frac{D}2} \, \partial_0^2 \Bigl(\frac{i\delta\!
\Delta_C}{ \Delta x^{D-2}} \Bigr) = O(D\!-\!4) \; .
\end{equation}
The final answer for this sub-section is therefore,
\begin{equation}
-i M^2_{\delta C}(x;x') = \frac{i e^2 H^2}{2^6 \pi^2} \, a^4 \delta^4(x 
\!-\! x') + O(D\!-\!4) \; . \label{dCfin}
\end{equation}

\subsection{Renormalization}

As we have explained, no other terms can give nonzero contributions for
$D\!=\!4$. The total for $-i M^2(x;x')$ at one loop order is therefore the
sum of (\ref{4fin}), (\ref{cffin}), (\ref{dAfin}), (\ref{dBfin}) and 
(\ref{dCfin}),
\begin{eqnarray}
\lefteqn{-i M^2(x;x') = } \nonumber \\
& & \frac{i e^2 \mu^{D-4}}{4 \pi^{\frac{D}2}} 
\frac{\Gamma(\frac{D}2\!-\!1)}{(D\!-\!3)(D\!-\!4)} \, a a' \partial^2 
\delta^D(x\!-\!x') -\frac{i e^2 \mu^{D-4}}{8 \pi^{\frac{D}2}} 
\Biggl\{\frac{\Gamma(\frac{D}2\!+\!1)}{(D\!-\!3) (D\!-\!4)} \nonumber \\
& & + 2 \ln\Bigl(\frac{H}{2 \mu}\Bigr) \!-\! \frac12 \Biggr\} H^2 a^4 
\delta^D(x\!-\!x') \!-\! \frac{e^2}{2^8 \pi^4} \, a a' \partial^6 \Bigl\{ 
\ln^2(\mu^2 \Delta x^2) \!-\! 2 \ln(\mu^2 \Delta x^2) \Bigr\} \nonumber \\
& & - \frac{e^2 H^4}{2^6 \pi^4} \, (a a')^3 \Biggl\{ \partial_0^2 
\Bigl[ \ln^2\Bigl(2^{-2} H^2 \Delta x^2\Bigr) \!-\!  \ln\Bigl(2^{-2} H^2 
\Delta x^2\Bigr) \Bigr] \nonumber \\
& & \hspace{1.7cm} + \frac{\partial^2}2 \Bigl[3 \ln^2\Bigl(2^{-2} H^2
\Delta x^2\Bigr) \!-\! 5 \ln\Bigl(2^{-2} H^2 \Delta x^2\Bigr) \Bigr] \Biggr\} 
+ O(D\!-\!4) \; .\qquad \label{M2simp}
\end{eqnarray}
By simply deleting (\ref{dAfin}) we can also obtain the one loop
self-mass-squared for a conformally coupled scalar,
\begin{eqnarray}
\lefteqn{-i {\cal M}^2(x;x') = \frac{i e^2 \mu^{D-4}}{4 \pi^{\frac{D}2}} 
\frac{\Gamma(\frac{D}2\!-\!1)}{(D\!-\!3)(D\!-\!4)} \, a a' \partial^2 
\delta^D(x\!-\!x') } \nonumber \\
& & \hspace{2.5cm} - \frac{e^2}{2^8 \pi^4} \, a a' \partial^6 \Bigl\{ 
\ln^2(\mu^2 \Delta x^2) \!-\! 2 \ln(\mu^2 \Delta x^2) \Bigr\} 
+ O(D\!-\!4) \; . \qquad \label{M2conf}
\end{eqnarray}

The relevant counterterms (\ref{ctm}) give rise to a diagram with the topology
depicted in Fig.~3. The contribution it makes is,
\begin{eqnarray}
\lefteqn{-i M^2_{\mbox{\tiny ctm}}(x;x') \!=\! i \delta Z_2 \partial_{\mu} 
\Bigl( \sqrt{-g} g^{\mu\nu} \partial_{\nu} \Bigr) \delta^D(x \!-\! x') \!-\! 
i \delta\xi R \sqrt{-g} \, \delta^D(x \!-\! x') \; , \qquad } \\
& & = i \delta Z_2 (a a')^{\frac{D}2-1} \partial^2 \delta^D\!(x\!-\!x') 
\nonumber \\
& & \hspace{3cm} - i \Bigl[\delta \xi \!-\! \frac14 \Bigl(\frac{D \!-\!2}{D
\!-\!1}\Bigr) \delta Z_2 \Bigr] (D\!-\!1) D H^2 a^D 
\delta^D\!(x\!-\!x') . \qquad \label{M2ctm}
\end{eqnarray}
\begin{center}
\begin{picture}(300,50)(0,0)
\DashArrowLine(150,25)(90,25){2}
\DashArrowLine(210,25)(150,25){2}
\Vertex(150,25){3}
\Text(151,25)[]{\LARGE $\times$}
\Text(150,15)[b]{$x$}
\end{picture}
\\ {\rm Fig.~3: Contribution from counterterms.}
\end{center}
By comparing (\ref{M2simp}) with (\ref{M2ctm}) we see that the simplest
choice of counterterms is,
\begin{eqnarray}
\delta Z_2 & = & -\frac{e^2 \mu^{D-4}}{4 \pi^{\frac{D}2}} \frac{\Gamma(
\frac{D}2 \!-\!1)}{(D\!-\!3) (D\!-\!4)} \; , i\label{bestZ} \\
\delta \xi & = & -\frac{e^2 \mu^{D-4}}{8 \pi^{\frac{D}2}(D\!-\!1) D} 
\Biggl\{\frac{3 \Gamma(\frac{D}2 \!+\!1) }{(D\!-\!3) (D\!-\!4)} \!+\! 2 
\ln\Bigl(\frac{H}{2 \mu}\Bigr) \!-\! \frac12 \Biggr\} \; .\qquad \label{bestxi}
\end{eqnarray}
This gives the following fully renormalized result,
\begin{eqnarray}
\lefteqn{-i M^2_{\mbox{\tiny ren}}(x;x') = -\frac{i e^2}{8 \pi^2} \, a a'
\ln(a a') \partial^2 \delta^4(x\!-\!x') \!+\! \frac{i e^2 H^2}{4 \pi^2} \, 
a^4 \ln(a) \delta^4(x\!-\!x') } \nonumber \\
& & \hspace{-.5cm} - \frac{e^2}{2^8 \pi^4} \, a a' \partial^6 \Bigl\{ 
\ln^2(\mu^2 \Delta x^2) \!-\! 2 \ln(\mu^2 \Delta x^2) \Bigr\} \!-\! 
\frac{e^2 H^4}{2^6 \pi^4} \, (a a')^3 \Biggl\{ \partial_0^2 \Bigl[ 
\ln^2\Bigl(2^{-2} H^2 \Delta x^2\Bigr) \nonumber \\
& & \hspace{.5cm} - \ln\Bigl(2^{-2} H^2 \Delta x^2\Bigr) \Bigr] \!+\! 
\frac{\partial^2}2 \Bigl[3 \ln^2\Bigl(2^{-2} H^2\Delta x^2\Bigr) \!-\! 5
 \ln\Bigl(2^{-2} H^2 \Delta x^2\Bigr) \Bigr] \Biggr\} \; . \qquad \label{M2ren}
\end{eqnarray}

To renormalize the self-mass-squared of a conformally coupled scalar 
(\ref{M2conf}) the best choice of counterterms would be,
\begin{equation}
\delta Z_2 \Bigl\vert_{\rm conf} = 4 \Bigl(\frac{D\!-\!1}{D\!-\!2}\Bigr) 
\delta \xi \Bigl\vert_{\rm conf} = -\frac{e^2 \mu^{D-4}}{4 \pi^{\frac{D}2}} 
\frac{\Gamma( \frac{D}2 \!-\!1)}{(D\!-\!3) (D\!-\!4)} \; . \label{cfctm}
\end{equation}
With this choice we get the following renormalized self-mass-squared,
\begin{eqnarray}
\lefteqn{-i {\cal M}^2_{\mbox{\tiny ren}}(x;x') = -\frac{i e^2 }{8 \pi^2} 
\, a a' \ln(a a') \partial^2 \delta^4(x\!-\!x') } \nonumber \\
& & \hspace{4cm} - \frac{e^2}{2^8 \pi^4} \, a a' \partial^6 \Bigl\{ 
\ln^2(\mu^2 \Delta x^2) \!-\! 2 \ln(\mu^2 \Delta x^2) \Bigr\} 
\; . \qquad \label{cfren}
\end{eqnarray}
The vastly greater complexity of the minimally coupled result (\ref{M2ren})
derives from inflationary particle production, which the conformally coupled
scalar does not experience. It is also worth noting that the conformally
coupled result can be put in a manifestly de Sitter invariant form using the
de Sitter length function $y(x;x')$ (\ref{y}) and the conformal d'Alembertian,
\begin{equation}
\mathcal{D}_{\rm cf} \equiv \partial_{\mu} \Bigl( \sqrt{-g} g^{\mu\nu} 
\partial_{\nu} \Bigr) - \frac16 R \sqrt{-g} = a \partial^2 a \; .
\end{equation}
The resulting expression is,
\begin{equation}
-i {\cal M}^2_{\mbox{\tiny ren}}(x;x') = -\frac{e^2 H^2}{32 \pi^4} \, 
\mathcal{D}_{\rm cf} \mathcal{D}_{\rm cf}' \Biggl\{ \frac{\ln[y(x;x') 
\mu^2/H^2]}{y(x;x')} \Biggr\} \; . \qquad \label{dSinv}
\end{equation}
Note its similarity to one loop scalar self-mass-squared found for a
Yukawa-coupled scalar \cite{DW}. The manifest de Sitter invariance of this
expression proves that our use of a noninvariant gauge poses no problem at
least at one loop order.

\section{$M^2(x;x')$ in Allen-Jacobson Gauge}

One can see from expression (\ref{EM}) for the electromagnetic Lagrangian
that a possible invariant gauge fixing term is,
\begin{equation}
\mathcal{L}_{\rm AJ} = -\frac12 \Bigl( g^{\mu\nu} A_{\mu ; \nu} \Bigr)^2
\sqrt{-g} = -\frac12 a^{D-4} \Bigl(\eta^{\mu\nu} A_{\mu , \nu} - (D-2) H a
A_0\Bigr)^2 \; .
\end{equation}
In this gauge the photon propagator obeys,
\begin{equation}
\sqrt{-g} \Bigl[(D^2)^{\mu}_{~\rho} - R^{\mu}_{~\rho}\Bigr] i\Bigl[{}^{\rho}
\Delta_{\nu}\Bigr](x;x') = \delta^{\mu}_{~\nu} i\delta^D(x-x') \; ,
\label{inveqn}
\end{equation}
where the contravariant vector covariant derivative operator is,
\begin{eqnarray}
\lefteqn{(D^2)^{\mu}_{~\rho} \equiv g^{\alpha\beta} \Biggl\{\delta^{\mu}_{~
\rho} \partial_{\alpha} \partial_{\beta} + \Gamma^{\mu}_{~\alpha\rho}
\partial_{\beta} + \Gamma^{\mu}_{~\beta\rho} \partial_{\alpha} } \nonumber \\
& & \hspace{3cm} - \delta^{\mu}_{~\rho} \Gamma^{\nu}_{~\alpha\beta}
\partial_{\nu} + \Gamma^{\mu}_{~\alpha\rho , \beta} + \Gamma^{\mu}_{~\alpha\nu}
\Gamma^{\nu}_{~\beta\rho} - \Gamma^{\mu}_{~\nu\rho} \Gamma^{\nu}_{~\alpha\beta}
\Biggr\} . \qquad
\end{eqnarray}
Of course equation (\ref{inveqn}) is generally covariant in addition to being
de Sitter invariant, but we will only solve it for the special case of de
Sitter background.

Allen and Jacobson expressed their result for the solution of (\ref{inveqn})
in terms of scalar functions multiplying, respectively, the parallel transport
matrix and the product of two gradients of $\ell(x;x')$ \cite{AJ}. For our
purposes it is more effective to express the same result in terms of
the length function $y(x;x')$ --- which was defined in (\ref{y}) --- and its
derivatives,\footnote{One can show \cite{RPW} that the parallel transport 
matrix and the product of the two gradients take the following form in terms 
of $y(x;x')$ and its derivatives,
\begin{eqnarray}
\Bigl[\mbox{}_{\mu} g_{\nu}\Bigr](x;x') & = & -\frac1{2 H^2} \,
\frac{\partial^2 y}{\partial x^{\mu} \partial x^{\prime \nu}} - \frac1{2 H^2 (4
\!-\!y)} \, \frac{\partial y}{\partial x^{\mu}} \frac{\partial y}{\partial
x^{\prime \nu}} \; , \nonumber \\
\Bigl[\mbox{}_{\mu} n\Bigr](x;x') \Bigl[n_{\nu}\Bigr](x;x') & = & \frac1{H^2 y
(4 \!-\!y)} \, \frac{\partial y}{\partial x^{\mu}} \frac{\partial y}{\partial
x^{\prime \nu}} \; . \nonumber
\end{eqnarray}
Hence the functions $\alpha$ and $\beta$ of Allen and Jacobson \cite{AJ}
relate to our functions $B(y)$ and $C(y)$ as follows,
\begin{eqnarray}
\alpha & = & -2 H^2 B(y) \; , \nonumber \\
\beta & = & - y H^2 B(y) + y (4\!-\!y) H^2 C(y) \; . \nonumber
\end{eqnarray}}
\begin{equation}
i\Bigl[{}_{\mu} \Delta_{\nu}^{\rm AJ}\Bigr](x;x') = B(y) \frac{\partial^2 y}{
\partial x^{\mu} \partial x^{\prime \nu}} + C(y) \frac{\partial y}{\partial
x^{\mu}} \frac{\partial y}{\partial x^{\prime \nu}} \; , \label{AJprop}
\end{equation}
It is straightforward to compute various derivatives of $y(x;x')$ in
conformal coordinates,
\begin{eqnarray}
\frac{\partial y}{\partial x^{\mu}} & = & H a \Bigl(y \delta^0_{\mu} + 2 a' H
\Delta x_{\mu}\Bigr) \; , \\
\frac{\partial y}{\partial x^{\prime \nu}} & = & H a' \Bigl(y \delta^0_{\nu}
- 2 a H \Delta x_{\nu}\Bigr) \; , \\
\frac{\partial^2 y}{\partial x^{\mu} \partial x^{\prime \nu}} & = & H^2 a a'
\Bigl(y \delta^0_{\mu} \delta^0_{\nu} - 2 \delta^0_{\mu} a H \Delta x_{\nu}
+ 2 a' H \Delta x_{\mu} \delta^0_{\nu} - 2 \eta_{\mu\nu}\Bigr) \; ,
\end{eqnarray}
where the contravariant interval is $\Delta x_{\mu} \equiv \eta_{\mu\nu}
(x^{\nu} - x^{\prime \nu})$. One can also establish the following useful
results in any coordinate system,
\begin{eqnarray}
g^{\mu\nu}(x) \frac{\partial y}{\partial x^{\mu}} \frac{\partial y}{\partial
x^{\nu}} & = & H^2 \Bigl(4 y - y^2\Bigr) = g^{\mu\nu}(x') \frac{\partial y}{
\partial x^{\prime \mu}} \frac{\partial y}{\partial x^{\prime \nu}} \; , \\
g^{\mu\nu}(x) \frac{\partial y}{\partial x^{\nu}} \frac{\partial^2 y}{
\partial x^{\mu} \partial x^{\prime \sigma}} & = & H^2 (2-y) \frac{\partial y}{
\partial x^{\prime \sigma}} \; , \\
g^{\rho\sigma}(x') \frac{\partial y}{\partial x^{\prime \sigma}}
\frac{\partial^2 y}{\partial x^{\mu} \partial x^{\prime \rho}} & = & H^2 (2-y)
\frac{\partial y}{\partial x^{\mu}} \; , \\
g^{\mu\nu}(x) \frac{\partial^2 y}{\partial x^{\mu} \partial x^{\prime \rho}}
\frac{\partial^2 y}{\partial x^{\nu} \partial x^{\prime \sigma}} & = & 4 H^4
g_{\rho\sigma}(x') - H^2 \frac{\partial y}{\partial x^{\prime \rho}}
\frac{\partial y}{\partial x^{\prime \sigma}} \; , \\
g^{\rho\sigma}(x') \frac{\partial^2 y}{\partial x^{\mu}\partial x^{\prime \rho}}
\frac{\partial^2 y}{\partial x^{\nu} \partial x^{\prime \sigma}} & = & 4 H^4
g_{\mu\nu}(x) - H^2 \frac{\partial y}{\partial x^{\mu}} \frac{\partial y}{
\partial x^{\nu}} \; , \\
\frac{\partial^2 y}{D x^{\mu} D x^{\nu}} & = & H^2 (2-y) g_{\mu\nu}(x) \; , \\
\frac{\partial^3 y}{D x^{\mu} D x^{\nu} \partial x^{\prime \rho}} & = & -H^2
g_{\mu\nu}(x) \frac{\partial y}{\partial x^{\prime \rho}} \; , \\
D^2 \frac{\partial^2 y}{\partial x^{\mu} \partial x^{\prime \nu}} & = & -H^2
\frac{\partial^2 y}{\partial x^{\mu} \partial x^{\prime \nu}} \; .
\end{eqnarray}

Substituting (\ref{AJprop}) into (\ref{inveqn}) and making use of the various
identities gives the following coupled, ordinary differential equations for
$B(y)$ and $C(y)$,
\begin{eqnarray}
(4 y - y^2) B'' + D (2-y) B' - D B + (4 - 2y) C & = & 0 \; , \label{B''} \\
(4 y - y^2) C'' + (D+4) (2-y) C' - 2 D C - 2 B' & = & 0 \; . \label{C''}
\end{eqnarray}
This coupling is one indication of the substantially greater complication of
working with a de Sitter invariant gauge. We solve this system the same way
Allen and Jacobson solved their analogous system \cite{AJ}, by partially
decoupling through the change of variables $F(y) \equiv B'(y) - C(y)$,
\begin{eqnarray}
& & (4 y - y^2) F'' + (D+2) (2-y) F' - 2 (D-1) F = 0 \; , \label{Feqn} \\
& & (4 y - y^2) B'' + (D+2) (2-y) B' - D B  = (4 - 2y) F \; . \label{BFeqn}
\end{eqnarray}
One then substitutes the general solution for $F(y)$ into the second 
equation.\footnote{The two linearly independent solutions to the $F$ equation 
(\ref{Feqn}) are,
\begin{eqnarray}
\mbox{}_2F_1\Bigl(D\!-\!1,2;\frac{D}2 \!+\! 1; 1\!-\!\frac{y}4\Bigr) 
\qquad {\rm and} \qquad
\mbox{}_2F_1\Bigl(D\!-\!1,2;\frac{D}2 \!+\! 1; \frac{y}4\Bigr) \; . \nonumber
\end{eqnarray}
The two homogeneous equations to the $B$ equation (\ref{BFeqn}) are,
\begin{eqnarray}
\mbox{}_2F_1\Bigl(D,1;\frac{D}2 \!+\! 1; 1\!-\!\frac{y}4\Bigr) \qquad {\rm and}
\qquad \mbox{}_2F_1\Bigl(D,1;\frac{D}2 \!+\! 1;\frac{y}4\Bigr) \; . \nonumber
\end{eqnarray}}
The various integration constants are chosen to enforce the delta function
singularity at $y \!=\! 0$ and analyticity at $y \!=\! 4$.
In the end one finds,
\begin{eqnarray}
\lefteqn{B(y) = \frac{H^{D-4}}{(4\pi)^{\frac{D}2}} \frac{\Gamma(D\!-\!3)}{
\Gamma(\frac{D}2)} \Biggl\{-\Bigl(\frac{D\!-\!3}2\Bigr) \; 
\mbox{}_2F_1\Bigl(D\!-\!2,1;\frac{D}2;1\!-\!\frac{y}4\Bigr) } \nonumber \\
& & \hspace{.5cm} + \Bigl[\frac12 - \Bigl(\frac{D\!-\!2}{D}\Bigr) \Bigl(\psi(D)
\!-\! \psi(1)\Bigr) \Bigr] \, \mbox{}_2F_1\Bigl(D,1;\frac{D}2\!+\!1;\frac{y}4
\Bigr) \nonumber \\
& & \hspace{.5cm} + \frac{(D\!-\!3)}{(4y \!-\! y^2)^{\frac{D}2}} \!\int_0^y 
\! dy' (4y'\!-\! y^{\prime2})^{\frac{D}2-1} \!\! \int_{y'}^4 \!\! dy''
\mbox{}_2F_1\Bigl(D\!-\!2,1;\frac{D}2;1\!-\!\frac{y''}4\Bigr) \Biggr\} ,
\qquad \label{BHyper} \\
\lefteqn{C(y) = \frac{H^{D-4}}{(4\pi)^{\frac{D}2}} \frac{\Gamma(D\!-\!3)}{
\Gamma(\frac{D}2)} \frac{\partial}{\partial y} \Biggl\{\!
\Bigl[\frac12 \!-\! \Bigl(\frac{D\!-\!2}{D}\Bigr) \! \Bigl(\psi(D) \!-\! 
\psi(1)\Bigr) \Bigr] \, \mbox{}_2F_1\Bigl(D,1;\frac{D}2\!+\!1;\frac{y}4\Bigr) }
\nonumber \\
& & \hspace{.5cm} + \frac{(D\!-\!3)}{(4y \!-\! y^2)^{\frac{D}2}} \!\int_0^y
\!dy' (4y'\!-\! y^{\prime2})^{\frac{D}2-1} \!\!\int_{y'}^4 \!\! dy''
\mbox{}_2F_1\Bigl(D\!-\!2,1; \frac{D}2;1\!-\! \frac{y''}4\Bigr) \Biggr\} ,
\qquad \label{CHyper} \\
\lefteqn{F(y) = \frac{H^{D-4}}{(4\pi)^{\frac{D}2}} \frac{\Gamma(D\!-\!3)}{
\Gamma(\frac{D}2)} \times -\Bigl(\frac{D\!-\!3}2\Bigr) \frac{\partial}{
\partial y} \; \mbox{}_2F_1\Bigl(D\!-\!2,1;\frac{D}2; 1\!-\!\frac{y}4\Bigr)
\; , } \label{FHyper} \\
& & \hspace{.3cm} = -\frac{H^{D-4}}{(4\pi)^{\frac{D}2}} \frac{\Gamma(D\!-\!1)}{
\Gamma(\frac{D}2)} \frac{\partial}{\partial y} \Biggl\{\frac1{(4y\!-\!y^2)^{
\frac{D}2-1}} \int_y^4 dy' (4y' \!-\! y^{\prime 2})^{\frac{D}2 -2} \Biggr\} , 
\label{F2}
\end{eqnarray}
where $\psi(z) \equiv \Gamma'(z)/\Gamma(z)$.

Simplicity is sometimes a matter of taste but it is difficult to imagine any
criterion by which the propagator in this gauge is simpler than the one we
used in the previous section. Where the de Sitter invariant formalism has a 
clear advantage is in taking derivatives and contracting indicies. This is 
because derivatives of a function of $y(x;x')$, such as $B(y)$ or $C(y)$, 
produce ordinary derivatives of these functions with respect to $y$ times the 
same basis tensors as in the photon propagator. Invariant contractions of 
these basis tensors always reduce to functions or $y$. For example, one can 
easily compute the divergence of the photon propagator on $x^{\nu}$,
\begin{eqnarray}
\lefteqn{\partial_{\nu} \Biggl( \sqrt{-g(x)} g^{\mu\nu}(x)
i\Bigl[{}_{\mu} \Delta_{\rho}^{\rm AJ}\Bigr](x;x') \Biggr) = H^2 a^D \frac{
\partial y}{\partial x^{\prime \rho}} } \nonumber \\
& & \hspace{2.5cm} \times \Biggl\{(2\!-\!y) B' \!-\! D B \!+\! (4y\!-\!y^2) C'
\!+\! (D\!+\!1) (2\!-\!y) C\Biggr\} . \qquad
\end{eqnarray}
By first substituting $C(y) \!=\! B'(y) \!-\! F(y)$ and then using the $B$
equation (\ref{BFeqn}) we can actually express this in terms of $F(y)$ and
its first derivative,
\begin{eqnarray}
\lefteqn{(2\!-\!y) B' \!-\! D B \!+\! (4y\!-\!y^2) C' \!+\! (D\!+\!1) 
(2\!-\!y) C } \nonumber \\
& & \hspace{-.5cm} = (4y \!-\! y^2) B'' \!+\! (D\!+\!2) (2\!-\!y) B' \!-\! D B 
\!-\!  (4y\!-\!y^2) F' \!-\! (D\!+\!1) (2\!-\!y) F \; , \qquad \\
& & \hspace{-.5cm} = -(4y \!-\! y^2) F' \!-\! (D\!-\!1) (2\!-\!y) F \; .
\end{eqnarray}
Of course a similar result pertains for the divergence on $x^{\prime \sigma}$,
\begin{eqnarray}
\lefteqn{\partial^{\prime}_{\sigma} \Biggl( \sqrt{-g(x')} g^{\rho\sigma}(x')
i\Bigl[{}_{\mu} \Delta_{\rho}^{\rm AJ}\Bigr](x;x') \Biggr) } \nonumber \\
& & \hspace{3cm} = -H^2 a^{\prime D} \frac{\partial y}{\partial x^{\mu}} 
\Biggl\{ (4y \!-\! y^2) F' \!+\! (D\!-\!1) (2\!-\!y) F \Biggr\} . \qquad
\end{eqnarray}
When both divergences are taken the result can be reduced to a scalar, then
the second derivatives of $F(y)$ can be removed using the $F$ equation 
(\ref{Feqn}),
\begin{eqnarray}
\lefteqn{\partial_{\nu} \partial^{\prime}_{\sigma} \Biggl( \sqrt{-g(x)}
g^{\mu\nu}(x) \sqrt{-g(x')} g^{\rho\sigma}(x') i\Bigl[{}_{\mu} \Delta_{\rho}^{
\rm AJ} \Bigr](x;x') \Biggr) = - a^D \delta^D(x \!-\!x') } \nonumber \\
& & - (D\!-\!1) H^4 (a a')^D \Biggl\{ (2\!-\!y) (4y\!-\!y^2) F' \!+\!
(D\!-\!1) (2\!-\!y)^2 F \!+\! 4 F \Biggr\} . \qquad
\end{eqnarray}

To take advantage of these relations we re-express the 3-point contribution
(\ref{inv3}) as follows,
\begin{eqnarray}
\lefteqn{-i M^2_{\mbox{\tiny 3AJ}}(x;x') = -4 e^2 \sqrt{-g} g^{\mu\nu}
\sqrt{-g'} g^{\prime \rho\sigma} i\Bigl[\mbox{}_{\mu} \Delta^{\rm AJ}_{\rho}
\Bigr](x;x') \partial_{\nu} \partial_{\sigma}' i\Delta_A(x;x') } \nonumber \\
& & \hspace{2cm} - 2 e^2 \partial_{\nu} \Biggl( \sqrt{-g} g^{\mu\nu} i\Bigl[
\mbox{}_{\mu} \Delta^{\rm AJ}_{\rho}\Bigr](x;x') \Biggr) \sqrt{-g'} g^{\prime 
\rho\sigma} \partial_{\sigma}' i\Delta_A(x;x') \nonumber \\
& & \hspace{2cm} - 2 e^2 \partial_{\sigma}' \Biggl( \sqrt{-g'} g^{\prime 
\rho\sigma} i\Bigl[\mbox{}_{\mu} \Delta^{\rm AJ}_{\rho}\Bigr](x;x') \Biggr) 
\sqrt{-g} g^{\mu\nu} \partial_{\nu} i\Delta_A(x;x') \nonumber \\
& & \hspace{2cm} - e^2 \partial_{\nu} \partial_{\sigma}' \Biggl( \sqrt{-g} 
g^{\mu\nu} \sqrt{-g'} g^{\prime \rho\sigma} i\Bigl[\mbox{}_{\mu} 
\Delta^{\rm AJ}_{\rho} \Bigr](x;x') \Biggr) i\Delta_A(x;x') \; . \qquad 
\label{3AJ}
\end{eqnarray}
Because the scalar propagator is not quite de Sitter invariant we must also
extract its de Sitter breaking term,
\begin{equation}
i\Delta_A(x;x') \equiv A(y) + k \ln(a a') \qquad {\rm where} \qquad
k \equiv \frac{H^{D-2}}{(4\pi)^{\frac{D}2}} \frac{\Gamma(D\!-\!1)}{
\Gamma(\frac{D}2)} \; .
\end{equation}
The necessary first and second derivatives are,
\begin{eqnarray}
\partial_{\nu} i\Delta_A(x;x') & = & \frac{\partial y}{\partial x^{\nu}} A'(y)
+ k H a \delta_{\nu}^0 , \; \\
\partial^{\prime}_{\sigma} i\Delta_A(x;x') & = & \frac{\partial y}{\partial x^{
\prime \sigma}} A'(y) + k H a' \delta_{\sigma}^0 \; , \\
\partial_{\nu} \partial^{\prime}_{\sigma} i\Delta_A(x;x') & = &
\frac{\partial^2 y}{\partial x^{\nu} \partial x^{\prime \sigma}} A'(y) +
\frac{\partial y}{\partial x^{\nu}} \frac{\partial y}{\partial x^{\prime
\sigma}} A''(y) \nonumber \\
& & \hspace{3cm} + \delta^0_{\nu} \delta^0_{\sigma} a^{2-D} i \delta^D(x \!-\!
x') \; . \qquad
\end{eqnarray}
Note also that the function $A(y)$ obeys the inhomogeneous equation,
\begin{equation}
(4y \!-\! y^2) A'' + D (2 \!-\! y) A' - (D \!-\! 1) k = 0 \; . \label{A''}
\end{equation}

With these identities, and some of the previous ones for the photon propagator
and contractions, the first term of (\ref{3AJ}) reduces to,
\begin{eqnarray}
\lefteqn{-i M^2_{\mbox{\tiny 3AJ}_1}(x;x') \equiv -4 e^2 \sqrt{-g} g^{\mu\nu}
\sqrt{-g'} g^{\prime \rho\sigma} i\Bigl[\mbox{}_{\mu} \Delta^{\rm AJ}_{\rho}
\Bigr](x;x') \partial_{\nu} \partial_{\sigma}' i\Delta_A(x;x') \; ,} \\
& & \hspace{-.5cm} = -i 8 e^2 H^2 B a^D \delta^D(x\!-\!x') - 4 e^2 H^4 
(a a')^D \Biggl\{ \Bigl[ 4D \!-\! (4y \!-\! y^2)\Bigr] A' B \nonumber \\
& & \hspace{1cm} + (2 \!-\! y) (4 y \!-\! y^2) A' C + (2 \!-\! y) (4 y \!-\! 
y^2) A'' B + (4 y \!-\! y^2)^2 A'' C \Biggr\} , \qquad \\
& & \hspace{-.5cm} = -i 8 e^2 H^2 B a^D \delta^D(x\!-\!x') + 4 k (D\!-\!1) 
e^2 H^4 (a a')^D \Bigl[-(2\!-\!y) B \!-\! (4y \!-\! y^2) C\Bigr] \nonumber \\
& & \hspace{3cm} + 4 (D\!-\!1) e^2 H^4 (a a')^D (4 y \!-\! y^2) A' \Bigl[-B + 
(2\!-\! y) C \Bigr] \; . \label{phot1}
\end{eqnarray}
The second term of (\ref{3AJ}) gives,
\begin{eqnarray}
\lefteqn{-i M^2_{\mbox{\tiny 3AJ}_2}(x;x') \equiv - 2 e^2 \partial_{\nu} 
\Bigl( \sqrt{-g} g^{\mu\nu} i\Bigl[\mbox{}_{\mu} \Delta^{\rm AJ}_{\rho}
\Bigr](x;x') \Bigr) \sqrt{-g'} g^{\prime \rho\sigma} \partial_{\sigma}' 
i\Delta_A(x;x') , \qquad } \\
& & \hspace{2cm} = 2 e^2 H^4 (a a')^D \Bigl\{ (4 y \!-\! y^2) A' \!+\! k 
\Bigl[2 \!-\! y \!-\! 2 \frac{a}{a'}\Bigr] \Bigr\} \nonumber \\
& & \hspace{6.2cm} \times \Bigl\{(4 y \!-\! y^2) F' \!+\! (D\!-\!1) (2 \!-\! y) 
F\Bigr\} . \qquad
\end{eqnarray}
The third term differs only by interchanging $x^{\mu}$ and $x^{\prime \mu}$, 
so the two sum to,
\begin{eqnarray}
\lefteqn{-i M^2_{\mbox{\tiny 3AJ}_{2+3}}(x;x') = 4 e^2 H^4 (a a')^D \Bigl\{
(4 y \!-\! y^2) A' \!-\! k \Bigl[y \!+\! a a' H^2 \Delta \eta^2 \Bigr]
\Bigr\} } \nonumber \\
& & \hspace{6cm} \times \Bigl\{(4 y \!-\! y^2) F' \!+\! (D\!-\!1) 
(2 \!-\! y) F\Bigr\} . \qquad\label{phot23}
\end{eqnarray}
The fourth term in (\ref{3AJ}) gives,
\begin{eqnarray}
\lefteqn{-i M^2_{\mbox{\tiny 3AJ}_4}(x;x') \equiv - e^2 \partial_{\nu} 
\partial_{\sigma}' \Biggl( \sqrt{-g} g^{\mu\nu} \sqrt{-g'} g^{\prime 
\rho\sigma} i\Bigl[\mbox{}_{\mu} \Delta^{\rm AJ}_{\rho} \Bigr](x;x') \Biggr) 
i\Delta_A(x;x') , \qquad } \\
& & = i e^2 \Bigl\{ A \!+\! 2 k \ln(a) \Bigr\} a^D 
\delta^D(x\!-\!x') \!+\! (D\!-\!1) e^2 H^4 (a a')^D \Bigl\{A \!+\! k 
\ln(a a')\Bigr\} \nonumber \\
& & \hspace{3.8cm} \times \Biggl\{ (2 \!-\! y) (4y \!-\! y^2) F' 
\!+\! (D\!-\!1) (2\!-\!y)^2 F \!+\! 4 F \Biggr\} . \qquad \label{phot4}
\end{eqnarray}
Finally, we should include the 4-point contribution,
\begin{eqnarray}
-i M^2_{\mbox{\tiny 4AJ}}(x;x') & \equiv & -i e^2 \sqrt{-g} g^{\mu\nu} i\Bigl[
\mbox{}_{\mu} \Delta^{\rm AJ}_{\nu}\Bigr](x;x') \, \delta^D(x-x') \; , \\
& = & i 2 D e^2 H^2 B a^D \delta^D(x\!-\!x') \; .
\end{eqnarray}

Up to this point we have simply exploited the differential equations 
(\ref{A''}) and (\ref{Feqn}-\ref{BFeqn}) that any solution for the scalar
and photon propagators must obey. Now we substitute the specific solutions 
for $A(y)$, $B(y)$, $C(y)$ and $F(y)$, and then expand the relevant 
combinations to include all singular and finite, nonzero contributions. The 
expansion of $A(y)$ can be read off from (\ref{DeltaA}) and has the nice 
property of terminating for $D\!=\!4$. The two combinations we require are,
\begin{eqnarray}
A(y) & = & \frac{H^{D-2}}{4 \pi^{\frac{D}2}} \Biggl\{ \frac{\Gamma(\frac{D}2
\!-\!1)}{y^{\frac{D}2\!-\!1}} \!-\! \frac12 \ln\Bigl(\frac{y}4\Bigr) \!-\!
\frac14 \!+\! O(D\!-\!4) \Biggr\} , \label{Aexp} \\
(4y \!-\! y^2) A'(y) & = & -\frac{H^{D-2}}{4 \pi^{\frac{D}2}} \Biggl\{
\frac{4 \Gamma(\frac{D}2)}{y^{\frac{D}2-1}} \!+\! \frac{(\frac{D}2 \!-\! 1)
\Gamma(\frac{D}2)}{y^{\frac{D}2-2}} \!-\! \frac{y}2 \!+\! O(D\!-\!4) \Biggr\} 
. \qquad \label{A'exp}
\end{eqnarray}
The function $F(y)$ also terminates in $D\!=\!4$,
\begin{eqnarray}
\lefteqn{F(y) = \frac{H^{D-4}}{16 \pi^{\frac{D}2}} \Biggl\{ \frac{2 \Gamma(
\frac{D}2)}{y^{\frac{D}2}} + \frac{(D\!-\!4) \Gamma(\frac{D}2)}{2 y^{\frac{D}2
-1}} + \frac1{2^{D-1}} \sum_{n=0}^{\infty} \Biggl[ (n\!+\!1) 
\frac{\Gamma(n\!+\! D\!-\!1)}{\Gamma(n\!+\!\frac{D}2\!+\!1)} 
\Bigl(\frac{y}4\Bigr)^n } \nonumber \\
& & \hspace{5cm} - \Bigl(n \!-\! \frac{D}2 \!+\! 3\Bigr) \frac{\Gamma(n\!+\!
\frac{D}2\!-\!2)}{(n\!+\!2)!} \Bigl(\frac{y}4\Bigr)^{n-\frac{D}2+2} \Biggr] 
\Biggr\} . \qquad 
\end{eqnarray}
Two combinations of it are needed,
\begin{eqnarray}
(4y \!-\! y^2) F' \!+\! (D\!-\!1) (2 \!-\! y) F \!=\! \frac{H^{D-4}}{16
\pi^{\frac{D}2}} \Biggl\{\! -\frac{4 \Gamma(\frac{D}2)}{y^{\frac{D}2}} \!-\!
\frac{\Gamma(\frac{D}2\!+\!1)}{y^{\frac{D}2-1}} \!+\! O(D\!-\!4)\! \Biggr\} , 
\quad \label{exp23} \\
(2\!-\!y) (4y \!-\! y^2) F' \!+\! (D\!-\!1) (2 \!-\! y)^2 F \!+\! 4 F \!=\! 
\frac{H^{D-4}}{16 \pi^{\frac{D}2}} \times \frac{\Gamma(D\!-\!1)}{2^{D-4}
\Gamma(\frac{D}2)} \; . \quad \label{exp4}
\end{eqnarray}
The last expression can be shown to be exact using (\ref{F2}).

Unlike $i\Delta_B(x;x')$ and $i\Delta_C(x;x')$, the expansions for $B(y)$ and 
$C(y)$ do not terminate in $D\!=\!4$ dimensions. (This is another indication 
of the complication associated with employing a de Sitter invariant gauge.) 
However, we can take $D\!=\!4$ in those terms which vanish rapidly enough at 
$y\!=\!0$ to make only finite contributions to the process under study. For 
the one loop self-mass-squared the relevant expansions are,
\begin{eqnarray}
B(y) & = & \frac{H^{D-4}}{16 \pi^{\frac{D}2}} \Biggl\{ -\frac{2 \Gamma(
\frac{D}2 \!-\! 1)}{y^{\frac{D}2-1}} + \Delta B(y) + O\Bigl( (D\!-\!4)
\ln(y)\Bigr) \Biggr\} , \\
C(y) & = & \frac{H^{D-4}}{16 \pi^{\frac{D}2}} \Biggl\{ -\frac{\Gamma(
\frac{D}2 \!-\! 1)}{2 y^{\frac{D}2-1}} + \Delta C(y) + O\Bigl( (D\!-\!4)
\ln(y)\Bigr) \Biggr\} , 
\end{eqnarray}
where the $D\!=\!4$ residual functions are,
\begin{eqnarray}
\Delta B(y) & \equiv & \Biggl[-\frac{\frac83}{(4\!-\!y)^2} - \frac{\frac43}{4
\!-\!y} \Biggl] \ln\Bigl(\frac{y}4\Bigr) - \frac{\frac23}{4\!-\!y} \; , \\
\Delta C(y) & \equiv & \Biggl[-\frac{\frac{16}3}{(4\!-\!y)^3} - \frac{\frac43}{
(4\!-\!y)^2} \Biggl] \ln\Bigl(\frac{y}4\Bigr) - \frac{\frac43}{(4\!-\!y)^2}
- \frac{\frac12}{4\!-\!y} \; .
\end{eqnarray}
The two combinations for which we require expansions are,
\begin{eqnarray}
-(2\!-\!y) B \!-\! (4y \!-\! y^2) C \!=\! \frac{H^{D-4}}{2^4 \pi^{\frac{D}2}} 
\Biggl\{\frac{4 \Gamma(\frac{D}2 \!-\! 1)}{y^{\frac{D}2-1}} \!+\! \frac{16 
\ln(\frac{y}4)}{(4 \!-\! y)^2} \!+\! \frac{(\frac43 \!+\! \frac23 y)}{4\!-\!y}
\Biggl\} , \label{exp1a} \\
-B \!+\! (2\!-\! y) C \!=\! \frac{H^{D-4}}{2^4 \pi^{\frac{D}2}} \Biggl\{
\frac{\Gamma(\frac{D}2 \!-\! 1)}{y^{\frac{D}2 -1}} \!+\! \frac{\frac{32}3 
\ln(\frac{y}4)}{(4 \!-\! y)^3} \!+\! \frac{(4 \!-\! \frac13 y)}{(4 \!-\! y)^2} 
\Biggl\} . \label{exp1b}
\end{eqnarray}

Substituting (\ref{exp1a}-\ref{exp1b}) and (\ref{A'exp}) in (\ref{phot1})
gives,
\begin{eqnarray}
\lefteqn{-i M^2_{\mbox{\tiny 3AJ}_1}(x;x') \!=\! -i 8 e^2 H^2 B a^D 
\delta^D(x\!-\!x') } \nonumber \\
& & \hspace{1.5cm} + \frac{e^2 H^{2D-2}}{2^4 \pi^D} (a a')^D \Biggl\{ -\frac{8 
(D\!-\!1) \Gamma^2(\frac{D}2)}{(D\!-\!2) y^{D-2}} \!-\! \frac8{y (4\!-\!y)} 
\ln\Bigl(\frac{y}4\Bigr)\Biggr\} .\qquad
\end{eqnarray}
Although the de Sitter breaking part of the scalar propagator contributes to
this expression, it does so in a de Sitter invariant fashion. Explicit breaking
of de Sitter invariance appears when we substitute (\ref{exp23}) and 
(\ref{A'exp}) in (\ref{phot23}),
\begin{eqnarray}
\lefteqn{-i M^2_{\mbox{\tiny 3AJ}_{2+3}}(x;x') = \frac{e^2 H^{2D-2}}{2^4 \pi^D}
(a a')^D } \nonumber \\
& & \hspace{1cm} \times \Biggl\{\frac{16 \Gamma^2(\frac{D}2)}{y^{D-1}} \!+\! 
\frac{4 (D\!-\!1) \Gamma^2(\frac{D}2)}{y^{D-2}} + \frac2{y} + a a' H^2 \Delta 
\eta^2 \Bigl[ \frac2{y^2} \!+\! \frac1{y} \Bigr] \Biggr\} . \qquad
\end{eqnarray}
Note that the de Sitter breaking terms are finite so they require no 
noninvariant counterterms. Because (\ref{exp4}) is nonsingular it makes a 
similarly finite (and similarly non-de Sitter invariant) contribution when 
substituted with (\ref{Aexp}) in (\ref{phot4}),
\begin{eqnarray}
\lefteqn{-i M^2_{\mbox{\tiny 3AJ}_4}(x;x') = 
\frac{e^2 H^6}{2^4 \pi^4} (a a')^4 \Biggl\{\frac{3}2 \frac1{y} \!-\!
\frac34 \ln\Bigl(\frac{\sqrt{e}}4 H^2 \Delta x^2\Bigr) \Biggl\} } \nonumber \\
& & \hspace{1.5cm} + \frac{i e^2 H^{D-2}}{(4\pi)^{\frac{D}2}} 
\frac{\Gamma(D\!-\!1)}{\Gamma(\frac{D}2)} \Bigl\{-\pi \cot\Bigl(\frac{\pi}2 D
\Bigr) + 2 \ln(a)\Bigr\} a^D \delta^D(x\!-\!x') . \qquad
\end{eqnarray} 
We can now sum the various terms to obtain the regulated but unrenormalized
result,
\begin{eqnarray}
\lefteqn{-i M^2_{\mbox{\tiny AJ}}(x;x') = 
+i e^2 H^2 2 (D\!-\!4) B a^D \delta^D(x \!-\! x') } \nonumber \\
& & \hspace{.5cm} + \frac{i e^2 H^{D-2}}{(4\pi)^{\frac{D}2}} \frac{\Gamma(D\!-
\!1)}{\Gamma(\frac{D}2)} \Bigl\{-\pi \cot\Bigl(\frac{\pi}2 D\Bigr) + 
2 \ln(a)\Bigr\} a^D \delta^D(x\!-\!x') \nonumber \\
& & \hspace{.5cm} + \frac{e^2 H^{2D-2}}{2^4 \pi^D} (a a')^D \Biggl\{\frac{16 
\Gamma^2(\frac{D}2)}{y^{D-1}} \!+\! \frac{4 (D\!-\!4) (D\!-\!1) 
\Gamma^2(\frac{D}2)}{(D\!-\!2) y^{D-2}} \!+\! \frac1{2 y} \nonumber \\
& & \hspace{1.5cm} - \frac{8}{y (4\!-\!y)} \ln\Bigl(\frac{y}4\Bigr) \!+\! a a' 
H^2 \Delta \eta^2 \Bigl[ \frac2{y^2} \!+\! \frac1{y} \Bigr] \!-\! \frac34 
\ln\Bigl(\frac{\sqrt{e}}4 H^2 \Delta x^2\Bigr) \Biggl\} . \qquad
\end{eqnarray}
Only two terms require partial integration,
\begin{eqnarray}
\lefteqn{\frac{e^2 H^{2D-2}}{2^4 \pi^D} (a a')^D \times \frac{16 \Gamma^2(
\frac{D}2)}{y^{D-1}} = \frac{e^2}{4 \pi^D} a a' \frac{(D\!-\!2)^2 \Gamma^2(
\frac{D}2 \!-\! 1)}{\Delta x^{2D-2}} \; , } \\
& & \hspace{2cm} = \frac{i e^2 \mu^{D-4}}{4 \pi^2} \frac{\Gamma(\frac{D}2 
\!-\! 1)}{(D\!-\! 3)(D\!-\!4)} a a' \partial^2 \delta^D(x\!-\!x') \nonumber \\
& & \hspace{4cm} - \frac{e^2}{2^5 \pi^4} a a' \partial^4 \Bigl(\frac{\ln(\mu^2 
\Delta x^2)}{\Delta x^2}\Bigr) + O(D\!-\!4) \; , \qquad \\
\lefteqn{\frac{e^2 H^{2D-2}}{2^4 \pi^D} (a a')^D \times \frac{4 (D\!-\!4)
(D\!-\!1) \Gamma^2(\frac{D}2)}{(D\!-\!2) y^{D-2}} } \nonumber \\
& & \hspace{2cm} = \frac{e^2 H^2}{4 \pi^D} (a a')^2 \frac{(D\!-\!4)(D\!-\!1) 
\Gamma^2(\frac{D}2)}{(D \!-\! 2) \Delta x^{2D-4}} \; , \\
& & \hspace{2cm} = \frac{3 i e^2 H^2}{4 \pi^2} a^4 \delta^4(x \!-\! x') + 
O(D\!-\!4) \; .
\end{eqnarray}
It is also worth noting that the $(D\!-\!4)$ times the coincidence limit of
$B(y)$ is finite in $D\!=\!4$ dimensions,
\begin{equation}
i e^2 H^2 2 (D\!-\!4) B a^D \delta^D(x\!-\!x') = -\frac{i e^2 H^2}{8 \pi^2}
a^4 \delta^4(x\!-\!x') + O(D\!-\!4) \; .
\end{equation}

The simplest choice of renormalization constants seems to be,
\begin{eqnarray}
\delta Z^{\mbox{\tiny AJ}} & = & -\frac{e^2 \mu^{D-4}}{4 \pi^{\frac{D}2}} 
\frac{\Gamma(\frac{D}2 \!-\!1)}{(D\!-\!3) (D\!-\!4)} \; , \\
\delta \xi^{\mbox{\tiny AJ}} & = & \frac14 \Bigl(\frac{D\!-\!2}{D\!-\!1}\Bigr) 
\delta Z_{\mbox{\tiny AJ}} - \frac{e^2 H^{D-4}}{(4 \pi)^{\frac{D}2}(D\!-\!1)D} 
\frac{\Gamma(D\!-\!1)}{\Gamma(\frac{D}2)} \Bigl\{\pi \cot\Bigl(\frac{\pi}2 D
\Bigr) \!-\! 5\Bigr\} \; . \qquad
\end{eqnarray}
Note that the field strength renormalization in Allen-Jacobson gauge is 
exactly the same (\ref{bestZ}) that we found for the simple gauge of the 
previous section. The conformal term in Allen-Jacobson gauge is different 
from what we found (\ref{bestxi}) for the simple gauge, but only by finite 
terms. So the divergence structure seems to be the same.

With these conventions the fully renormalized self-mass-squared is,
\begin{eqnarray}
\lefteqn{-i M^2_{\mbox{\tiny AJ}_{\rm ren}}(x;x') = -\frac{i e^2}{8 \pi^2} \, 
a a'\ln(a a') \partial^2 \delta^4(x\!-\!x') \!+\! \frac{i e^2 H^2}{4 \pi^2} \, 
a^4 \ln(a) \delta^4(x\!-\!x') } \nonumber \\
& & - \frac{e^2}{2^8 \pi^4} \, a a' \partial^6 \Bigl\{ \ln^2(\mu^2 \Delta x^2) 
\!-\! 2 \ln(\mu^2 \Delta x^2) \Bigr\} + \frac{e^2 H^6}{16 \pi^4} (a a')^4 
\Biggl\{\frac1{2y} \nonumber \\
& & \hspace{1.8cm} - \frac8{y (4\!-\!y)} \ln\Bigl(\frac{y}4\Bigr) \!+\! a a' 
H^2 \Delta \eta^2 \Bigl[ \frac2{y^2} \!+\! \frac1{y} \Bigr] \!-\! \frac34
\ln\Bigl(\frac{\sqrt{e}}4 H^2 \Delta x^2\Bigr) \Biggl\} . \qquad \label{AJren}
\end{eqnarray}
The first three terms agree with what we got (\ref{M2ren}) in the simple gauge.
The remaining terms show some similarities but they do not agree. Although 
1PI functions can differ, off-shell, when computed in different gauges 
\cite{Jackiw}, there seems to be a clear disagreement. To see this note
that, for large $y(x;x')$, the Allen-Jacobson self-mass-squared (\ref{AJren}) 
differs by a factor of $\frac3{16} e^2 H^4 (a a')^4 i \Delta_A(x;x')$. There
is no chance that such a term gives zero when integrated against a wave
function, which is how one goes ``on-shell'' in position space. In fact the 
integral is {\it singular}! 

The singularity is not present in (\ref{M2ren}) so we suspect a problem with 
the homogeneous terms in the Allen-Jacobson propagator. Specifically, it may
not be correct to enforce analyticity at $y\!=\!4$. To clarify the situation we 
will use $i\Bigl[\mbox{}_{\mu} \Delta^{\rm AJ}_{\rho}\Bigr](x;x')$ to compute 
the self-mass-squared for a conformally coupled scalar whose propagator is,
\begin{equation}
i\Delta_{\rm cf}(x;x') = \frac{H^{D-2}}{4 \pi^{\frac{D}2}}
\frac{\Gamma(\frac{D}2\!-\!1)}{y^{\frac{D}2-1}} \equiv A_{\rm cf}(y) \; .
\end{equation}
This field does not experience inflationary particle production, and our
result (\ref{cfren}) for its self-mass-squared in the simple gauge is a
trivial conformal rescaling of the flat space result. That seems to be 
correct on physical grounds so we can use it to check
$i\Bigl[\mbox{}_{\mu} \Delta^{\rm AJ}_{\rho}\Bigr](x;x')$.

Of course the photon structure is not changed, and most of our previous 
analysis is still valid. What changes is that the prefactor $k$ of the de 
Sitter breaking term is zero and, rather than (\ref{A''}), derivatives of 
the function $A_{\rm cf}$ are all proportional to one another,
\begin{equation}
A_{\rm cf}'(y) = -\Bigl(\frac{D}2 \!-\!1\Bigr) \frac{A_{\rm cf}}{y} \qquad 
{\rm and} \qquad A_{\rm cf}''(y) = \Bigl(\frac{D}2 \!-\!1\Bigr) \frac{D}2 
\frac{A_{\rm cf}}{y^2} \; .
\end{equation}
We require one new combination of the photon functions,
\begin{equation}
B \!-\! (4\!-\!y) C = \frac{H^{D-4}}{16 \pi^{\frac{D}2}} \Biggl\{
\frac{\frac83}{(4\!-\!y)^2} \ln\Bigl(\frac{y}4\Bigr) \!+\! 
\frac{\frac23}{4 \!-\! y} \!+\! O(D\!-\!4) \Biggr\} . \label{exp1c}
\end{equation}
Using these results and (\ref{exp1a}) one finds,
\begin{eqnarray}
\lefteqn{-i \mathcal{M}^2_{\mbox{\tiny 3AJ}_1}(x;x') \equiv -4 e^2 \sqrt{-g} 
g^{\mu\nu} \sqrt{-g'} g^{\prime \rho\sigma} i\Bigl[\mbox{}_{\mu} \Delta^{\rm 
AJ}_{\rho}\Bigr](x;x') \partial_{\nu} \partial_{\sigma}' i\Delta_{\rm cf}(x;x')
\; , } \\
& & = -i 8 e^2 H^2 B a^D \delta^D(x\!-\!x') - 4 e^2 H^4 (a a')^D
\Biggl\{ \Bigl[ 4D \!-\! (4y \!-\! y^2)\Bigr] A_{\rm cf}' B \nonumber \\
& & \hspace{.5cm} + (2 \!-\! y) (4 y \!-\! y^2) A_{\rm cf}' C + (2 \!-\! y) 
(4 y \!-\!  y^2) A_{\rm cf}'' B + (4 y \!-\! y^2)^2 A_{\rm cf}'' C \Biggr\} , 
\qquad \\
& & = -i 8 e^2 H^2 B a^D \delta^D(x\!-\!x') \!+\! 2 (D\!-\!2)
e^2 H^4 (a a')^D A_{\rm cf} \nonumber \\
& & \hspace{1cm} \times \Biggl\{2(D\!-\!1) \Bigl[B \!-\! (4\!-\!y) C\Bigr]
\!+\! \Bigl(\frac{D}2 \!-\! 1\Bigr) \Bigl[ (2\!-\!y) B \!+\! (4y \!-\! y^2) C 
\Bigr] \Biggr\} , \qquad \label{key1} \\
& & = -i 8 e^2 H^2 B a^D \delta^D(x\!-\!x') \nonumber \\
& & \hspace{3cm} + 
\frac{e^2 H^{2D-2}}{16 \pi^D} (a a')^D \Biggl\{ -\frac{4 \Gamma^2(
\frac{D}2)}{y^{D-2}} \!+\! \frac23 \frac1{y} \!+\! O(D\!-\!4) \Biggr\} . 
\qquad \label{cfph1}
\end{eqnarray}

The divergence on $x^{\nu}$ gives,
\begin{eqnarray}
\lefteqn{-i \mathcal{M}^2_{\mbox{\tiny 3AJ}_2}(x;x') \equiv - 2 e^2 
\partial_{\nu} \Biggl( \sqrt{-g} g^{\mu\nu} i\Bigl[\mbox{}_{\mu} 
\Delta^{\rm AJ}_{\rho} \Bigr](x;x') \Biggr) \sqrt{-g'} g^{\prime \rho\sigma} 
\partial_{\sigma}' i\Delta_{\rm cf}(x;x') } \nonumber \\
& & = (D\!-\!2) e^2 H^4 (a a')^D (4 \!-\! y) A_{\rm cf} \Bigl\{- (4 y \!-\! 
y^2) F' \!-\! (D\!-\!1) (2 \!-\! y) F\Bigr\} . \qquad
\end{eqnarray}
This is symmetric under interchange of $x^{\mu}$ and $x^{\prime \mu}$ so one
gets the same from the divergence on $x^{\prime \sigma}$, and the two 
divergence terms sum to,
\begin{eqnarray}
\lefteqn{-i \mathcal{M}^2_{\mbox{\tiny 3AJ}_{2+3}}(x;x') 
= 2 (D\!-\!2) e^2 H^4 (a a')^D (4 \!-\! y) A_{\rm cf} } \nonumber \\
& & \hspace{5cm} \times \Bigl\{- (4 y \!-\! y^2) F' \!-\! (D\!-\!1) 
(2 \!-\! y) F\Bigr\} . \qquad \label{key23} \\
& & = \frac{e^2 H^{2D-2}}{16 \pi^D} \Biggl\{ \frac{16 \Gamma^2(\frac{D}2)}{
y^{D-1}} + \frac{2(D\!-\!2) \Gamma^2(\frac{D}2)}{y^{D-2}} \!-\! \frac2{y}
\!+\! O(D\!-\!4) \Biggr\} . \qquad \label{cfph23}
\end{eqnarray}
Because the coincidence limit of $A_{\rm cf}(y)$ vanishes in dimensional
regularization we obtain no local contribution from the last term in
(\ref{3AJ}),
\begin{eqnarray}
\lefteqn{-i \mathcal{M}^2_{\mbox{\tiny 3AJ}_4}(x;x') \equiv -e^2 \partial_{\nu} 
\partial_{\sigma}' \Biggl( \sqrt{-g} g^{\mu\nu} \sqrt{-g'} g^{\prime 
\rho\sigma} i\Bigl[\mbox{}_{\mu} \Delta^{\rm AJ}_{\rho} \Bigr](x;x') \Biggr) 
i\Delta_{\rm cf}(x;x') \; , } \nonumber \\
& & \hspace{-.5cm} = (D\!-\!1) e^2 H^4 (a a')^D A_{\rm cf} \Biggl\{ (2\!-\!y) 
(4y \!-\! y^2) F' \!+\! (D\!-\!1) (2 \!-\!y)^2 F \!+\! 4F \Biggr\} , \qquad 
\label{key4} \\
& & \hspace{-.5cm} = \frac{e^2 H^6}{16 \pi^4} \Biggl\{ \frac32 \frac1{y} 
+ O(D\!-\!4) \Biggr\} . \label{cfph4}
\end{eqnarray}
Because it does not involve the scalar, the 4-point contribution is unchanged,
\begin{eqnarray}
-i \mathcal{M}^2_{\mbox{\tiny 4AJ}}(x;x') & \equiv & -i e^2 \sqrt{-g} 
g^{\mu\nu} i\Bigl[\mbox{}_{\mu} \Delta^{\rm AJ}_{\nu}\Bigr](x;x') \, 
\delta^D(x-x') \; , \\ & = & i 2 D e^2 H^2 B a^D \delta^D(x\!-\!x') \; .
\end{eqnarray}

It remains to sum the various contributions and isolate the divergences,
\begin{eqnarray}
\lefteqn{-i \mathcal{M}^2_{\mbox{\tiny AJ}}(x;x') = i 2 (D \!-\! 4) e^2 H^2 
B a^D \delta^D(x\!-\!x') } \nonumber \\
& & + \frac{e^2 H^{2D-2}}{16 \pi^D} (a a')^D \Biggl\{\frac{16 \Gamma^2(
\frac{D}2)}{y^{D-1}} \!+\! \frac{2(D\!-\!4) \Gamma^2(\frac{D}2)}{y^{D-2}} 
\!+\! \frac1{6 y} \!+\! O(D\!-\!4) \Biggr\} , \qquad \\
& & = \frac{i e^2 \mu^{D-4}}{4 \pi^{\frac{D}2}} \frac{\Gamma(\frac{D}2\!-\!1)}{
(D\!-\!3) (D\!-\!4)} \, a a' \partial^2 \delta^D(x\!-\!x') + \frac{i e^2 H^2}{
8 \pi^2} \, a^4 \delta^4(x\!-\!x') \nonumber \\
& & -\frac{e^2}{2^8 \pi^4} a a' \partial^6 \Bigl\{\ln^2(\mu^2 \Delta x^2) \!-\!
2 \ln(\mu^2 \Delta x^2)\Bigr\} \!+\! \frac{e^2 H^4}{96 \pi^4} \frac{(a a')^3}{
\Delta x^2} + O(D\!-\!4) \; . \qquad
\end{eqnarray}
The best choice of counterterms would seem to be,
\begin{eqnarray}
\delta Z_2^{\mbox{\tiny AJ}} \Bigl\vert_{\rm conf} & = & -\frac{e^2 \mu^{D-4}}{
4 \pi^{\frac{D}2}} \frac{\Gamma( \frac{D}2 \!-\!1)}{(D\!-\!3) (D\!-\!4)} \; ,\\
\delta \xi^{\mbox{\tiny AJ}} \Bigl\vert_{\rm conf} & = &
\frac{e^2 \mu^{D-4}}{4 \pi^{\frac{D}2} (D\!-\!1)D} \Biggl\{-
\frac{\Gamma( \frac{D}2\!+\!1)}{(D\!-\!3) (D\!-\!4)} + \frac12 \Biggr\} .
\end{eqnarray}
This choice gives the following renormalized self-mass-squared,
\begin{eqnarray}
\lefteqn{-i {\cal M}^2_{\mbox{\tiny AJ}_{\rm ren}}(x;x') = -\frac{i e^2 }{8 
\pi^2} \, a a' \ln(a a') \partial^2 \delta^4(x\!-\!x') } \nonumber \\
& & \hspace{2cm} - \frac{e^2}{2^8 \pi^4} \, a a' \partial^6 \Bigl\{ 
\ln^2(\mu^2 \Delta x^2) \!-\! 2 \ln(\mu^2 \Delta x^2) \Bigr\} \!+\! 
\frac{e^2 H^4}{96 \pi^4} \frac{(a a')^3}{\Delta x^2} \; . \qquad
\end{eqnarray}
The first two terms agree with what we found (\ref{cfren}) in the simple 
gauge, but the final term is again proportional to $(a a')$ times the
(conformal) scalar propagator. 

How did this happen? The conformal scalar propagator is of course correct,
as is the expression (\ref{3AJ}) for the 3-point contribution. It might seem
that a bewildering sequence of expansions have taken place but the astute 
reader will note that, except for local terms, the entire result is
proportional to the factor,
\begin{equation}
2(D\!-\!2) e^2 H^4 (a a')^D A_{\rm cf}(y) = \frac{e^2 H^{D+2}}{\pi^{\frac{D}2}}
\, (a a')^D \frac{\Gamma(\frac{D}2)}{y^{\frac{D}2-1}} \; ,
\end{equation}
times the sum of just three combinations of the functions $B(y)$, $C(y)$
and $F(y)$. The three key combinations can be read off from expressions
(\ref{key1}), (\ref{key23}) and (\ref{key4}),
\begin{eqnarray}
\lefteqn{2(D\!-\!1) \Bigl[B \!-\! (4\!-\!y) C\Bigr] + \Bigl(\frac{D}2\!-\!1
\Bigr) \Bigl[(2\!-\!y) B \!+\! (4y \!-\! y^2)C\Bigr] } \nonumber \\
& & \hspace{2.5cm} = \frac{H^{D-4}}{16 \pi^{\frac{D}2}} \Biggl\{- \frac{4 
\Gamma(\frac{D}2)}{y^{\frac{D}2-1}} + \frac23 + O(D\!-\!4) \Biggr\} , 
\label{BCE} \\
\lefteqn{-(4\!-\!y) \Bigl[(4y\!-\!y^2) F' \!+\! (D\!-\!1) (2\!-\!y) F\Bigr] }
\nonumber \\
& & \hspace{2.5cm} = \frac{H^{D-4}}{16 \pi^{\frac{D}2}} \Biggl\{\frac{16 
\Gamma(\frac{D}2)}{y^{\frac{D}2}} \!+\! \frac{2(D\!-\!2) \Gamma(\frac{D}2)}{
y^{\frac{D}2-1}} \!-\! 2 \!+\! O(D\!-\!4) \Biggr\} , \qquad \label{FE1} \\
\lefteqn{\frac12 \Bigl(\frac{D\!-\!1}{D\!-\!2}\Bigr) \Bigl[(2\!-\!y)
(4y\!-\!y^2) F' \!+\! (D\!-\!1) (2\!-\!y)^2 F \!+\! 4 F \Bigr] }
\nonumber \\
& & \hspace{2.5cm} = \frac{H^{D-4}}{16 \pi^{\frac{D}2}} \Biggl\{\frac32 +
O(D\!-\!4) \Biggr\} . \label{FE2}
\end{eqnarray}
The problem derived from the presence of the constant terms in each of these
three expansions. Note that it would not even be acceptable to have three
nonzero constants sum to zero because (\ref{FE1}) and (\ref{FE2}) contribute as
well to the minimally coupled self-mass-squared. Indeed, in that case the 
worst problem derives from the nonzero constant in just (\ref{FE2}). Whatever
we do must therefore expunge these constant terms.

What freedom do we have to alter $B(y)$, $C(y)$ and $F(y)$? First recall that
$C(y)$ is not independent, but instead obeys $C(y) \!=\! B'(y) \!-\! F(y)$. 
The functions $B(y)$ and $F(y)$ must of course satisfy the equations 
(\ref{Feqn}-\ref{BFeqn}) which define the photon propagator. However, we
suspect a homogeneous solution is missing. Because two of the three crucial
combinations involve only $F(y)$ we begin with it. The two independent 
solutions to the $F$ equation equation (\ref{Feqn}) and their $D\!=\!4$
limits are,
\begin{eqnarray}
F_1(y) & \equiv & \frac1{16} \mbox{}_2F_1\Bigl(D\!-\!1,2;\frac{D}2
\!+\!1;1\!-\! \frac{y}4\Bigr) \longrightarrow \frac1{y^2} \; , \\
F_2(y) & \equiv & \frac1{16} \mbox{}_2F_1\Bigl(D\!-\!1,2;\frac{D}2
\!+\!1;\frac{y}4 \Bigr) \longrightarrow \frac1{(4\!-\!y)^2} \; .
\end{eqnarray}
In $D\!=\!4$ these induce the following dependence in $B(y)$ through its
equation (\ref{BFeqn}),
\begin{eqnarray}
\delta B_1(y) &=& -\frac1{y} - \Biggl[\frac{\frac43}{(4\!-\!y)^2} \!+\!
\frac{\frac23}{4\!-\!y} \Biggr] \ln\Bigl(\frac{y}4\Bigr) - \frac{\frac13}{
4\!-\!y} \; , \\
\delta B_2(y) &=& \frac1{4\!-\!y} + \Biggl[\frac{\frac43}{y^2} \!+\!
\frac{\frac23}{y} \Biggr] \ln\Bigl(1\!-\!\frac{y}4\Bigr) + \frac{\frac13}{y} 
\; .
\end{eqnarray}

The solution of Allen and Jacobson is entirely based upon $F_1(y)$ and
$\delta B_1(y)$. The other solution, $F_2(y)$ and $\delta B_2(y)$, was 
rejected on the grounds that its singularity at the antipodal point
(the pole at $y\!=\!4$) would correspond to a point source there. Although 
this seemed correct to us as well, the problem we have encountered with the 
self-mass-squared makes us suspect that, on the full de Sitter manifold, the
photon propagator should have an anti-source at the antipodal point 
$\overline{x}^{\mu}$,
\begin{equation}
\sqrt{-g} \Bigl[(D^2)^{\mu}_{~\rho} - R^{\mu}_{~\rho}\Bigr] i\Bigl[{}^{\rho}
\Delta_{\nu}\Bigr](x;x') = \delta^{\mu}_{~\nu} i \Bigl[\delta^D(x\!-\!x') -
\delta^D(\overline{x}\!-\! x') \Bigr] \; . \label{neweqn}
\end{equation}
We offer two arguments in support of this suggestion. First, the linearization
instability of the gauge invariant theory precludes solutions with nonzero
charge, and the only de Sitter invariant position at which to locate the
compensating anti-charge is the antipodal point. Second, {\it including an 
anti-source makes the problem go away}. The contributions of $-F_2(y)/8\pi^2$, 
and of the terms it induces in $B(y)$ and $C(y)$, to expressions 
(\ref{BCE}-\ref{FE2}) are,
\begin{eqnarray}
\lefteqn{-\frac1{8 \pi^2} \Biggl\{2(D\!-\!1) \Bigl[\delta B_2 \!-\! (4\!-\!y) 
\delta C_2\Bigr] + \Bigl(\frac{D}2\!-\!1 \Bigr) \Bigl[(2\!-\!y) \delta B_2
\!+\! (4y \!-\! y^2) \delta C_2 \Bigr] \Biggr\} } \nonumber \\
& & \hspace{1.5cm} = \frac{H^{D-4}}{16 \pi^{\frac{D}2}} \Biggl\{
-\frac8{4\!-\!y} \!-\! \frac{128}{y^3} \ln\Bigl(1\!-\!
\frac{y}4\Bigr) \!-\! \frac{32}{y^2} \!-\! \frac4{y} \!-\! \frac23
\!+\! O(D\!-\!4) \Biggr\} , \qquad \\
\lefteqn{-\frac1{8\pi^2} \times -(4\!-\!y) \Bigl[(4y\!-\!y^2) F_2' \!+\! 
(D\!-\!1) (2\!-\!y) F_2 \Bigr] } \nonumber \\
& & \hspace{4cm} = \frac{H^{D-4}}{16 \pi^{\frac{D}2}} \Biggl\{
\frac4{4\!-\!y} + 2 + O(D\!-\!4) \Biggr\} , \qquad  \\
\lefteqn{-\frac1{8 \pi^2} \times \frac12 \Bigl(\frac{D\!-\!1}{D\!-\!2}\Bigr) 
\Bigl[(2\!-\!y) (4y\!-\!y^2) F_2' \!+\! (D\!-\!1) (2\!-\!y)^2 F_2 \!+\! 4 F_2
\Bigr] } \nonumber \\
& & \hspace{4cm} = \frac{H^{D-4}}{16 \pi^{\frac{D}2}} \Biggl\{
-\frac32 + O(D\!-\!4) \Biggr\} .
\end{eqnarray}

A final point is that there is nothing to be gained from the homogeneous
solutions to the $B$ equation (\ref{BFeqn}),
\begin{eqnarray}
B_1(y) & \equiv & \frac1{16} \mbox{}_2F_1\Bigl(D,1;\frac{D}2\!+\!1;1
\!-\! \frac{y}4\Bigr) \longrightarrow \frac1{y^2} + \frac{\frac12}{y} \; , \\
B_2(y) & \equiv & \frac1{16} \mbox{}_2F_1\Bigl(D,1;\frac{D}2\!+\!1;
\frac{y}4 \Bigr) \longrightarrow \frac1{(4\!-\!y)^2} + \frac{\frac12}{4\!-\!y}
\; .
\end{eqnarray}
Of course we must reject $B_1(y)$ on account of its singularity at $y\!=\!0$.
Because we do not care overmuch what happens at the antipodal point we might
allow $B_2(y)$. However, this does not affect relations (\ref{FE1}-\ref{FE2})
because they involve only $F(y)$. It turns out that $B_2(y)$ also has no
effect upon (\ref{BCE}), so there is nothing to be gained by adding it.

\section{Discussion}

We have used two different gauges to compute the renormalized 
self-mass-squared for a massless, minimally coupled scalar and for a 
massless, conformally coupled scalar. In the simplest gauge \cite{RPW}
our result for the minimally coupled scalar is,
\begin{eqnarray}
\lefteqn{M^2_{\mbox{\tiny ren}}(x;x') = \frac{e^2}{8 \pi^2} \, a a'
\ln(a a') \partial^2 \delta^4(x\!-\!x') \!-\! \frac{e^2 H^2}{4 \pi^2} \, 
a^4 \ln(a) \delta^4(x\!-\!x') } \nonumber \\
& & \hspace{-.5cm} - \frac{i e^2}{2^8 \pi^4} \, a a' \partial^6 \Bigl\{ 
\ln^2(\mu^2 \Delta x^2) \!-\! 2 \ln(\mu^2 \Delta x^2) \Bigr\} \!-\! 
\frac{i e^2 H^4}{2^6 \pi^4} \, (a a')^3 \Biggl\{ \partial_0^2 \Bigl[ 
\ln^2\Bigl(2^{-2} H^2 \Delta x^2\Bigr) \nonumber \\
& & \hspace{.5cm} - \ln\Bigl(2^{-2} H^2 \Delta x^2\Bigr) \Bigr] \!+\! 
\frac{\partial^2}2 \Bigl[3 \ln^2\Bigl(2^{-2} H^2\Delta x^2\Bigr) \!-\! 5
 \ln\Bigl(2^{-2} H^2 \Delta x^2\Bigr) \Bigr] \Biggr\} \; . \qquad \label{right}
\end{eqnarray}
Although this gauge breaks de Sitter invariance, it did not require
any noninvariant counterterms. Nor did it prevent the attainment of a
fully de Sitter invariant result for the conformally coupled scalar,
\begin{eqnarray}
\lefteqn{{\cal M}^2_{\mbox{\tiny ren}}(x;x') = \frac{e^2 }{8 \pi^2} 
\, a a' \ln(a a') \partial^2 \delta^4(x\!-\!x') } \nonumber \\
& & \hspace{4cm} - \frac{i e^2}{2^8 \pi^4} \, a a' \partial^6 \Bigl\{ 
\ln^2(\mu^2 \Delta x^2) \!-\! 2 \ln(\mu^2 \Delta x^2) \Bigr\} 
\; . \qquad
\end{eqnarray}

Working in a de Sitter invariant gauge \cite{AJ} was much more complicated.
Although all contractions and derivatives of de Sitter invariant objects
produce simple combinations of the propagator functions $B(y)$ and $C(y)$
and their derivatives, the propagator functions (\ref{BHyper}-\ref{CHyper})
are themselves so complicated as to preclude a simple analysis. It is 
particularly pointless to employ a de Sitter invariant gauge when the matter 
field in question necessarily breaks de Sitter invariance as does the 
minimally coupled scalar \cite{AF}.

There also seems to be a physical problem with the de Sitter invariant
propagator. This is evident from our renormalized results for the minimally 
coupled and conformally coupled scalars,
\begin{eqnarray}
\lefteqn{M^2_{\mbox{\tiny AJ}_{\rm ren}}(x;x') = \frac{e^2}{8 \pi^2} \, 
a a'\ln(a a') \partial^2 \delta^4(x\!-\!x') \!-\! \frac{e^2 H^2}{4 \pi^2} \, 
a^4 \ln(a) \delta^4(x\!-\!x') } \nonumber \\
& & - \frac{ie^2}{2^8 \pi^4} \, a a' \partial^6 \Bigl\{ \ln^2(\mu^2 \Delta x^2) 
\!-\! 2 \ln(\mu^2 \Delta x^2) \Bigr\} + \frac{i e^2 H^6}{16 \pi^4} (a a')^4 
\Biggl\{\frac1{2y} \nonumber \\
& & \hspace{1.5cm} - \frac8{y (4\!-\!y)} \ln\Bigl(\frac{y}4\Bigr) \!+\! a a' 
H^2 \Delta \eta^2 \Bigl[ \frac2{y^2} \!+\! \frac1{y} \Bigr] \!-\! \frac34
\ln\Bigl(\frac{\sqrt{e}}4 H^2 \Delta x^2\Bigr) \Biggl\} , \qquad \\
\lefteqn{{\cal M}^2_{\mbox{\tiny AJ}_{\rm ren}}(x;x') = \frac{e^2 }{8 
\pi^2} \, a a' \ln(a a') \partial^2 \delta^4(x\!-\!x') } \nonumber \\
& & \hspace{2cm} - \frac{i e^2}{2^8 \pi^4} \, a a' \partial^6 \Bigl\{ 
\ln^2(\mu^2 \Delta x^2) \!-\! 2 \ln(\mu^2 \Delta x^2) \Bigr\} \!+\! 
\frac{i e^2 H^4}{96 \pi^4} \frac{(a a')^3}{\Delta x^2} \; . \qquad
\end{eqnarray}
In each case the leading infrared contribution goes like $(a a')^4$ times
the relevant scalar propagator. Although off-shell quantities such as the
self-mass-squared can depend upon the choice of gauge \cite{Jackiw}, some 
of the extra terms we get for the de Sitter invariant gauge are actually 
singular on shell.

We suspect the problem is that the de Sitter invariant propagator should 
not be required to be analytic at the antipodal point. We demonstrated 
that the on-shell singularities disappear when the propagator is given 
an anti-source antipodal to the source. There may be a mathematical 
justification for this in the fact that the gauge invariant theory 
possesses a linearization instability which precludes solutions with 
nonzero net charge.

Of course this study was not undertaken to compare different gauges!
Our motivation was instead to learn how the scalar responds to the
dielectric medium of super-horizon scalars produced by inflation. It has 
previously been shown that this medium causes super-horizon photons to 
behave, in some ways, as if they have positive mass \cite{PW1,PTW1,PTW2,PW2}. 
Key questions we seek to answer are:
\begin{itemize}
\item{Does part of the scalar get ``eaten'' to make up the longitudinal
polarization of the massive photon as in the Higgs mechanism?}
\item{Does the one loop self-mass-squared enhance or retard further
particle production?}
\end{itemize}
Both questions can be answered by using the self-mass-squared (\ref{right})
to solve the linearized effective field equation (\ref{effeqn}) at one 
loop order. This will be undertaken in a subsequent paper.

A final application of this computation is to serve as ``data'' in
obtaining a stochastic formulation of SQED to sum the leading infrared
logarithms \cite{RPW2,TW}. Aside from its intrinsic interest, SQED 
provides a nearly perfect arena for the difficult task of generalizing
Starobinski\u{\i}'s techniques \cite{AAS,SY} from simple scalar models to
quantum gravity. It possesses derivative interactions and gauge constraints, 
it shows infrared logarithms {\it and} it is relatively tractable. This 
exercise has demonstrated the validity and viability of the Feynman rules in 
the noninvariant gauge. The stage has been set for the computations which 
are crucial for checking any stochastic formulation of SQED, the two loop 
vacuum expectation values,
\begin{equation}
\Bigl\langle \Omega \Bigl\vert \varphi^*(x) \varphi(x) \Bigr\vert \Omega
\Bigr\rangle \quad , \quad
\Bigl\langle \Omega \Bigl\vert F_{\rho\sigma}(x) F_{\mu\nu}(x) \Bigr\vert 
\Omega \Bigr\rangle \quad {\rm and} \quad 
\Bigl\langle \Omega \Bigl\vert T_{\mu\nu}(x) \Bigr\vert 
\Omega \Bigr\rangle \; .
\end{equation}

\vskip .3cm
\centerline{\bf Acknowledgments}

This work was partially supported by NSF grant PHY-0244714 and by the
Institute for Fundamental Theory at the University of Florida.

\end{document}